\documentclass[preprint, journal]{vgtc}                     

\onlineid{0}

\vgtccategory{Research}

\vgtcpapertype{Theoretical \& Empirical}

\title{Do LLMs Have Visualization Literacy? An Evaluation on Modified Visualizations to Test Generalization in Data Interpretation}

\author{%
  \authororcid{Jiayi Hong}{0000-0002-1332-5045},
  \authororcid{Christian Seto}{0009-0009-0808-2470},
  \authororcid{Arlen Fan}{0000-0002-8376-2881}, and 
  \authororcid{Ross Maciejewski}{0000-0001-8803-6355}
}

\authorfooter{
  \item
  	Jiayi Hong, Christian Seto, Arlen Fan, and Ross Maciejewski are with Arizona State University, USA.
  	E-mail: {\{jhong76\,$|$\,cseto\,$|$\,afan5\,$|$\,rmacieje\}}@asu.edu.
}

\abstract{%
  In this paper, we assess the visualization literacy of two prominent Large Language Models (LLMs): OpenAI's Generative Pretrained Transformers (GPT), the backend of ChatGPT, and Google's Gemini, previously known as Bard, to establish benchmarks for assessing their visualization capabilities. While LLMs have shown promise in generating chart descriptions, captions, and design suggestions, their potential for evaluating visualizations remains under-explored. Collecting data from humans for evaluations has been a bottleneck for visualization research in terms of both time and money, and if LLMs were able to serve, even in some limited role, as evaluators, they could be a significant resource. To investigate the feasibility of using LLMs in the visualization evaluation process, we explore the extent to which LLMs possess visualization literacy---a crucial factor for their effective utility in the field. We conducted a series of experiments using a modified 53-item Visualization Literacy Assessment Test (VLAT) for \gpt\ and \gemini. Our findings indicate that the LLMs we explored currently fail to achieve the same levels of visualization literacy when compared to data from the general public reported in VLAT, and LLMs heavily relied on their pre-existing knowledge to answer questions instead of utilizing the information provided by the visualization when answering questions. 
}

\keywords{Large Language Model, Visualization Literacy, \edit{Evaluation Study}}

\graphicspath{{figs/}{figures/}{pictures/}{images/}{./}} 

\usepackage{tabu}                      
\usepackage{booktabs}                  
\usepackage{lipsum}                    
\usepackage{mwe}                       

\usepackage{mathptmx}                  
\usepackage{multirow}
\usepackage{amsmath}
\usepackage{makecell}
\usepackage{siunitx}
\usepackage[svgnames]{xcolor}
\usepackage[normalem]{ulem}
\usepackage{wrapfig}
\usepackage{tabularx, booktabs}

\newcommand{\eg}{e.\,g.}
\newcommand{\ie}{i.\,e.}

\newcommand{\edit}[1]{\textcolor{black}{#1}}
\newcommand\redsout{\bgroup\markoverwith{\textcolor{red}{\rule[0.5ex]{2pt}{0.4pt}}}\ULon}
 \renewcommand{\redsout}[1]{}

\definecolor{better}{HTML}{43aa8b}
\definecolor{close}{HTML}{f9c74f}
\definecolor{worse}{HTML}{f94144}
\definecolor{gptColor}{HTML}{006E90}
\definecolor{geminiColor}{HTML}{F18F01}
\newcommand{\better}[1]{\textbf{\textcolor{better}{#1}}}
\newcommand{\close}[1]{\textcolor{close}{#1}}
\newcommand{\worse}[1]{\textcolor{worse}{#1}}

\newcommand{\tablefontsize}{\fontsize{7}{7}\selectfont}
\AtBeginEnvironment{tabu}{\tablefontsize}

\newcommand{\inlinetext}[1]{\raisebox{0pt}[0pt][0pt]{\raisebox{-0.6ex}{\includegraphics[height=2.4ex]{#1}}}}
\newcommand{\gpt}{\inlinetext{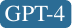}}
\newcommand{\gemini}{\inlinetext{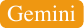}}

\begin{document}


\firstsection{Introduction}
\maketitle
\label{sec:intro}

As Generative Pretrained Transformers (GPT) and other Large Language Models (LLMs) continue to advance~\cite{Rahman:2023:CER}, their impact on the field of visualization is becoming increasingly profound. Previous research has explored implementing LLMs in diverse visualization tasks, \eg, visualization creation~\cite{Podo2024:V, DiBartolomeo:2023:AYSa}, data encoding~\cite{Shi:2023:NRC}, caption generation~\cite{Liew:2022:ULL}, design advice~\cite{Kim:2023:HGC}, and visualization education~\cite{Chen:2023:GCE}. Moreover, scientists have begun postulating the use of LLMs in evaluation studies (\eg, \cite{aubin:2024:LRT}). Such evaluations could also be useful for visualization sanity checks, providing feedback on visualization readability~\cite{Cabouat:2024:PPC}. Typically, evaluating visualizations incurs high costs related to human labor. If LLMs could support visualization experts in evaluating visual representations and visual systems to a similar level of proficiency as the general public, they could offer significant cost savings~\cite{Hamalainen:2023:ELL}. However, \edit{before experts can use LLMs to evaluate visualizations, it is crucial to ensure that these models possess adequate visualization literacy}. In this paper, we propose a \edit{template} to evaluate whether current LLMs have visualization literacy, which encompasses their ability to read, understand, and interpret visual representations, a crucial skill for evaluating visualizations. Such a \edit{template} is useful in determining when such technology has progressed to the point that we can consider its use in key tasks, such as evaluation.

To investigate the practicability of employing LLMs in the visualization assessment process, we systematically explored their visualization literacy. We assumed that an LLM is qualified to assist with visualization evaluation if it can meet the benchmark, indicating that its visualization literacy (VL) is comparable with that of the general public. Researchers have developed benchmarks for evaluating humans' VL such as the Visualization Literacy Assessment Test (VLAT)~\cite{Lee:2017:VDV} and its abridged version, Mini-VLAT~\cite{Pandey:2023:MiniVLAT} to gauge individuals' ability to interpret visual representations.
\edit{Yet, these static assessments cannot be directly adopted for LLMs, as these models may have already learned from such benchmarks. To this end, we developed a comprehensive evaluation template, incorporating a modified version of the 53-item VLAT} to evaluate the visualization literacy of LLMs, specifically focusing on \gpt\footnote{https://platform.openai.com/docs/models/gpt-4-and-gpt-4-turbo} and \gemini\footnote{https://ai.google.dev/models} in this work. This test covered 12 visualizations and 8 visual tasks. In a series of studies, we employed this modified test to evaluate the performance of LLMs with a primary focus on the accuracy rate as a metric. Additionally, we investigated the limitations of LLMs in reading and understanding visualizations, as well as the comparative costs between LLMs and humans. Our primary research questions are as follows:
\begin{itemize}
    \item \textbf{RQ1}: To what extent do LLMs have visualization literacy?
    \item \textbf{RQ2}: What are LLMs' limitations in interpreting visualizations?
    \item \textbf{RQ3}: What are the cost differences between LLMs and humans in interpreting visualizations and answering related questions?
\end{itemize}

We addressed \textbf{RQ1} by investigating the performances across different visualizations, tasks, and models. We also explored the extent to which LLMs' responses were influenced by their knowledge base versus their ability to extract and reason about data within visualizations. Moreover, we measured the costs associated with using LLMs in the experiments and compared them with \edit{the results of previously tested humans}. For \textbf{RQ2}, we examined the necessity of providing answer choices for LLMs and the impact of contextual information when it is kept in or removed from visualizations. Additionally, for \textbf{RQ3}, we measured the costs in terms of time and money we spent in the experiments to provide practitioners with an estimate of the costs for future usage.

Our results indicate that compared to humans, both \gpt\ and \gemini\ lack visualization literacy. Their performance varied across specific tasks and visualizations, and they heavily relied on their pre-existing knowledge base rather than the information present in the visualizations. Moreover, current LLMs seemed to perform poorly in open-ended visualization-related questions, and interestingly, decontextualization tended to enhance their performances, with \gpt\ showing more pronounced improvement compared to \gemini. Both LLMs are cost-effective compared with humans considering time and money, with \gemini\ showing greater cost-effectiveness than \gpt.

In summary, our contributions are
\begin{itemize}
    \item A template \edit{and methodology} for evaluating the visualization literacy of LLMs;
    \item A series of studies to examine the visualization literacy of LLMs;
    \item An exploration of the limitations of LLMs' in reading and interpreting visualizations;
    \item A discussion of insights on the future potential of leveraging LLMs in evaluating data visualizations.
\end{itemize}

\section{Related Work}
This work is grounded in theoretical foundations from visualization literacy, current applications of LLMs for visualization, and evaluation methods for LLMs.




\subsection{Visualization Literacy}

There has been a growing focus on exploring visualization literacy (VL) in society~\cite{Borner:2016:IAD} due to the widespread adoption of visualization across many domains~\cite{Bach:2021:Special}. Visualization literacy is vital, as it denotes the ability to read and interpret visual representations~\cite{Pandey:2023:MiniVLAT}. It enables individuals to comprehend unfamiliar charting techniques more easily~\cite{Borner:2019:Dvl}, which is essential in evaluating new visualizations.


To measure visualization literacy quantitatively, researchers have been creating evaluation methods. For example, Boy et al.~\cite{Boy:2014:Principled} spearheaded this endeavor by conceptualizing visualization literacy and introducing a method based on item response theory (IRT)~\cite{Baker:2001:BasicsIRT} for evaluating an individual's proficiency. 
Lee et al.~\cite{Lee:2017:VDV} developed and refined the Visualization Literacy Assessment Test (VLAT), which consists of 53 multiple-choice questions to assess visualization literacy among non-expert users. They highlighted a positive correlation between visualization literacy and the ability to interpret unfamiliar visualizations, suggesting its potential as a valuable tool for future research and educational applications. Expanding upon this groundwork, Pandey and Ottley~\cite{Pandey:2023:MiniVLAT} developed a condensed, 12-item form of the VLAT, known as the Mini-VLAT, to provide a quicker assessment tool for evaluating visualization literacy. 
Parallel to these developments, B{\"o}rner et al.~\cite{Borner:2019:Dvl} investigated the general public's visualization literacy by asking participants questions about their recognition and interpretation methods, shedding light on how people interact with visual information.

While previous studies have primarily focused on measuring humans' visualization literacy, there is a gap in our understanding of whether LLMs possess visualization literacy. This gap is particularly urgent and important to fill as LLMs are increasingly robust, and visualization practitioners are largely utilizing LLMs in completing visualization-related tasks~\cite{Vazquez:2024:ALR,Tian:2024:Chartgpt}. Our goal was to assess the level of visualization literacy possessed by LLMs, with the aim of better leveraging them in the future.
In our work, we define visualization literacy as \textbf{``the skill and ability to read, understand, and interpret information from visualization,''} loosely adapted from Lee et al.'s~\cite{Lee:2017:VDV} definition.

\subsection{Chart Question Answering}
\label{subsec:cqa}
Given our aim of utilizing LLMs to evaluate visualizations, it is crucial for these models to correctly answer visualization-related questions. This objective aligns directly with the field of chart question answering (CQA), where researchers are exploring diverse approaches to understand and interpret data presented in graphical formats.
DVQA~\cite{Kafle:2018:DVQA} introduced a synthetic dataset to test multiple aspects of bar chart understanding, illustrating the challenges faced by state-of-the-art visual question-answering (VQA) methods. The authors also presented novel deep neural network algorithms, demonstrating that tailored models are needed for specific types of visual data. ChartQA~\cite{Masry:2022:ChartQA} is another comprehensive dataset in the domain of CQA, featuring both human- and machine-generated question-answer pairs. This paper explored data augmentation through machine learning models to generate additional questions and hints at the possibility of using LLMs to create challenging datasets for CQA. Additionally, FigureQA~\cite{Kahou:2017:FigureQA} is a visual reasoning corpus based on synthetic and scientific-style figures, focusing on relational reasoning over plot elements. To answer these CQA questions, Kafle et al.~\cite{Kafle:2020:AnsweringPreFIL} introduced the PReFIL algorithm. This algorithm, tailored for complex CQA tasks, integrates parallel processing of questions and images to analyze both visual and text data from charts. Whereas most prior work in CQA focused on datasets and model benchmarking, their datasets were mostly real-world data. This setup can present challenges in determining whether models comprehensively understand the visualization or rely on their pre-existing knowledge to answer questions. Our work, in contrast, aims to conduct a comprehensive analysis to determine whether off-the-shelf LLMs possess visualization literacy.

\subsection{Evaluating LLMs}
As LLMs continue to show robustness and cost-efficiency, researchers are evaluating their performance across various domains to expand their applications~\cite{Chang:2024:SEL}.
One prominent area of study is programming, where researchers assessed LLMs' ability to generate code~\cite{chen:2021:evaluating, Liu:2023:REL, Vaithilingam:2022:EEE}. Additionally, LLMs have been tested in fields such as planning~\cite{Valmeekam:2023:OPA}, computational social science~\cite{Ziems:2024:LLM}, and psychology~\cite{Demszky:2023:ULL}.
In the human-computer interaction field, H\"am\"al\"ainen et al.~\cite{Hamalainen:2023:ELL} conducted studies to assess whether LLMs can simulate humans to provide open-ended text responses. The results indicated that they can provide human-like answers, but the answers would potentially be affected by malicious inputs. Chiang et al.~\cite{Chiang:2023:CLL} investigated the evaluation ability of LLMs to see whether they can replace human evaluation. These studies underscore the need for further research into the use of LLMs in various applications, including visualizations.

In the realm of visualization, researchers have been utilizing LLMs to create or enhance visualizations.
For example, Liew and Mueller~\cite{Liew:2022:ULL} have applied LLMs to generate meaningful captions for visualizations. Feng et al.~\cite{Feng:2023:PIP} developed a system to enable people to visually conduct prompt engineering in order to generate an ideal image from texts. ChartSpark~\cite{Xiao2023:ChartSpark} employs LLMs to integrate semantic context into charts through a novel text-to-image generative model. Such works demonstrate the power of LLMs for creating visualizations; however, evaluating the quality of the generated enhancements remains challenging~\cite{Rahman:2023:CER}. 
Kim et al.~\cite{Kim:2023:HGC} evaluated the capabilities of ChatGPT in giving advice on visualization design with the questions from the VisGuide forum. Chen et al.~\cite{Chen:2023:GCE} evaluated the performance of the GPT model in a data visualization course. In their experiments, they used SVG files and JavaScript code as input to test the GPT-4 model's ability to interpret visualizations. However, SVG files and JavaScript code are not commonly encountered in everyday contexts, such as news websites. While the results of their experiments suggest that GPT has the potential to complete diverse visualization tasks, GPT may simply be analyzing the provided code for the answer rather than truly interpreting, understanding, and reasoning about the charts. In contrast, our work focuses on systematically evaluating LLMs' visualization literacy using traditional PNG formats, which are more commonly encountered in real-world scenarios. Based on this work, Podo et al.~\cite{Podo2024:V} proposed a conceptual stack named EvaLLM to evaluate and interpret generative AI-based visualizations. 

Recent visualization researchers have also explored evaluating the visualization literacy of LLMs~\cite{Bendeck:2024:EEG, Li:2024:VLM}. However, these studies directly adopted questions and visualizations from VLAT, overlooking the prior knowledge inherent in LLMs. It is highly probable that the models had access to the visualizations, questions, and answers in their training data, potentially inflating their performance. Furthermore, they did not account for the influence of answer choices or the order in which options were presented~\cite{Pezeshkpour:2023:LLM}. Another recent work by Lo et al.~\cite{Lo:2024:HGB} investigated the capability of LLMs to identify misleading charts, but they also relied on the datasets directly collected from websites. To the best of our knowledge, no prior work has addressed a critical issue in LLMs' performance on visualization tasks: whether their responses are based on pre-existing knowledge or solely on the visualizations presented. This paper aims to explore that question.

\section{Pilot/exploratory Experiments}
Before performing a systematic investigation into LLMs' visualization literacy, we conducted a series of exploratory experiments to assess their current capabilities from two perspectives.

The first perspective involves extracting data from visualizations. We asked LLMs to generate data tables from various types of visualizations. We found that even with simple bar charts lacking value annotations, the models consistently made errors, failing to extract precise data in all five attempts, regardless of the presence of grid lines. Instead of accurately reading the lengths or positions of the charts, they heavily relied on the data labels within the visualizations, even when the data did not match the labels. Additionally, LLMs interpreted SVG files more accurately than \edit{PNG} files. However, since the SVG format is not widely used in everyday contexts, we focused on evaluating the LLMs' ability to read visualizations in the more commonly used \edit{PNG} format. The second perspective is about generating visualization code. We asked ChatGPT to generate Python code to recreate the provided charts and produce a new visualization based on the raw data. The results were suboptimal, as the model made errors in interpreting values and generating accurate bar heights. While the ChatGPT interface can display varying confidence levels in its responses, when using the ChatGPT API, the model consistently replies with the highest level of confidence, regardless of the accuracy of its output. We attached the tests results in \href{https://osf.io/wcb5g/}{\texttt{osf.io/wcb5g}}.

Given the results from our pilot experiments, we decided to investigate LLMs' visualization literacy by having them directly read visualizations in \edit{PNG} format, rather than transforming visualizations into text and then answering questions based on the text. Additionally, we removed data value labels from the charts to prevent LLMs from relying on the data labels.

\section{Assessing LLMs' Visualization Literacy}
\label{sec:assessingVL}
To address \textbf{RQ1}: \textit{To what extent do LLMs have visualization literacy?} we compared LLMs' performance with human results reported in previous work~\cite{Lee:2017:VDV}, which can provide a baseline for evaluation. Furthermore, we explored the effects of different conditions on LLMs' visualization literacy testing and investigated two additional questions:
\begin{itemize}
    \item \textbf{RQ1.1}: To what extent do LLMs' visualization literacy differ between models on different visualizations and tasks?
    \item \textbf{RQ1.2}: To what extent do LLMs answer questions with either their pre-existing knowledge or information in the visualization?
\end{itemize}
By examining LLMs' visualization literacy, we aimed to shed light on the current landscape of multimodal model usage in visualization comprehension and suggest future avenues for visualization evaluation. Our study scripts are available at \href{https://github.com/VADERASU/llm4viz-experiments}{\texttt{github.com/VADERASU/llm4viz-experiments}}.

\subsection{Experiments}
Although various LLMs were trained to read visualization-related texts, there are limited models openly accessible with image analysis capabilities. We utilized \gpt\ and \gemini\ in our study due to their widespread use and ease of access to their APIs for large-scale visualization analysis. Additionally, these models are multimodal, \ie, they allow multiple media inputs (text, images, and video). This allows us to feed PNG images of data visualizations directly to the LLMs as opposed to writing descriptions of the data visualization~\cite{Kim:2023:HGC} or providing the text of an SVG~\cite{Chen:2023:GCE}. To explore their visualization literacy, we adopted the assessment initially designed for evaluating human visualization literacy. We utilized the evaluation framework outlined in VLAT~\cite{Lee:2017:VDV} due to its comprehensive coverage of common visualization types and tasks and its validation by a broad audience. Additionally, in order to investigate whether LLMs base their answers on their knowledge or the information in the visualization, we conducted a separate experiment using the same questions but without visualizations.

\begin{figure*}
    \centering
    \includegraphics[width=\linewidth]{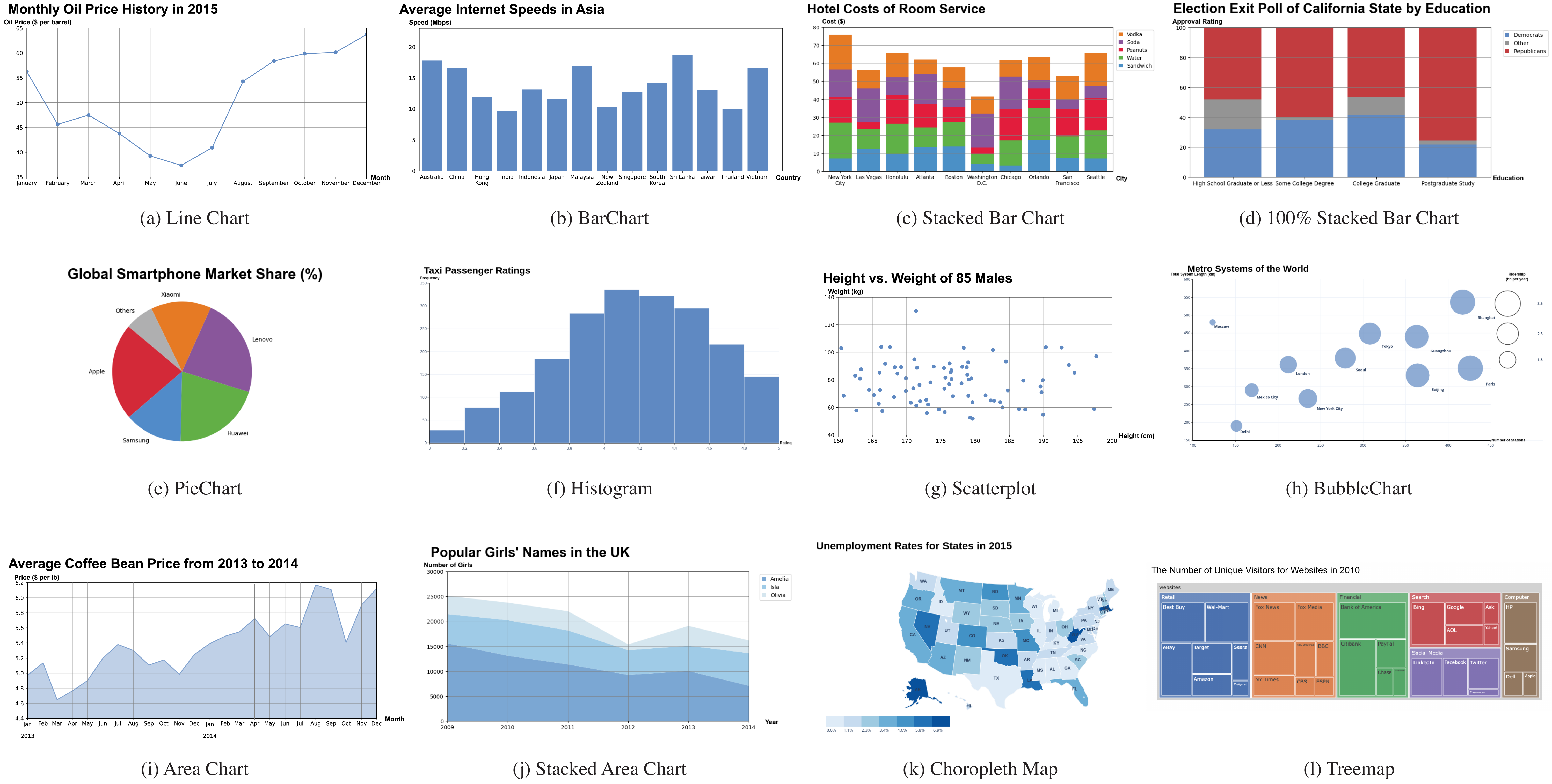}
    \caption{12 visualizations created based on VLAT examples and tested in our experiments.}
    \label{fig:visualizations}
\end{figure*}

\subsubsection{Experimental Design}
\label{subsec:vis_design}
To compare LLMs' performance with humans, as reported in the original VLAT paper, we initially intended to employ the exact same charts and questions used in VLAT. However, since VLAT was published in 2017, it is plausible that current LLMs have already encountered and incorporated these charts, questions, and answers into their training data. We conducted a pilot test to investigate \gpt's answers to the original VLAT visualization and questions. \gpt's responses should exhibit an equal probability distribution across all options if it were not trained. Yet, upon analysis of the results, we observed that for certain questions, \gpt\ consistently provided the same responses as those in VLAT, indicating potential exposure to the VLAT questions and answers or reliance on ground truth data from other sources during its training. As such, we decided to reproduce all the charts from VLAT to ensure that they were not part of LLMs' training corpus. We randomized values in our charts and strove to emulate the design style of the originals as closely as possible. Moreover, to enable LLMs to finish specific types of tasks like \textit{Find Correlation/Trends}, we manually adjusted the randomized data so that charts, \eg, the \textit{line chart}, can show trends. In addition, to fully distinguish the results from LLMs reading visualization from the results of VLAT, we updated the visualization, questions, and answers so that our answers were always different from the VLAT ones. The visualizations we created and used in the experiments are shown in \autoref{fig:visualizations}. After setting up the visualizations with questions and answers, we used \gpt's and \gemini's API for our tests. Before uploading the images online in January 2024, we double-checked the last knowledge update of \gpt, which occurred in April 2023. Since \gemini\ claimed that its knowledge is always being updated and refined once there is new knowledge, we did not publicly release our answers when conducting the studies. As such, \gemini\ is unlikely to have access to the answers to our modified assessment, and we can assume that our newly updated charts were not included in the model's training data while we were conducting our tests.

\subsubsection{Experiment 1: Evaluate LLMs' Visualization Literacy}
\label{subsec:vis+choices}
Using the visualizations outlined in Section~\ref{subsec:vis_design}, we then evaluated the visualization literacy of \gpt\ (model \texttt{gpt-4-vision-preview}) and \gemini\ (model \texttt{gemini-pro-vision}). Since we modified the data for all the charts, we updated the corresponding answers accordingly based on the experiment's questions. Initially, we tested \gpt\ following the options structure outlined in the VLAT paper, which includes an \textit{Omit} choice. However, as \gpt\ tended to answer \textit{Omit} when it was uncertain, we excluded the \textit{Omit} choice and prompted LLMs to make an educated guess if uncertain about the answer. To prevent LLMs from generating lengthy responses to questions, which can complicate the analysis process, we constrained its output to provide only the option letter, enabling a more focused analysis. In our pilot study, we experimented with various prompts and selected the most effective one for LLMs as follows:
\textit{``You are a helpful assistant for analyzing data visualizations. Please answer with the letter corresponding to the best option, or make a random guess if unsure. For example, if option (a) is correct, only reply with (a).''}
A total of 53 questions were investigated across various types of visualizations and tasks. Each question featured 2 (True/False questions) to 4 answer options (multiple-choice questions). As we found in our pilot study, the order of choices can affect the answer from LLMs. For each question, we thus conducted 120 tests with a counterbalanced order for all choices, meaning each combination of option orders was repeated between 120 / 4 = 30 and 120 / 2 = 60 times out of the total 120 tests. We shuffled all questions with a random seed to mitigate the effects of question order for the model. This results in a total of 53 $\times$ 120 = 6,360 trials. We recorded the time taken by LLMs to respond for further analysis. Given that the answer options for later questions may influence LLMs' responses to earlier questions, we conducted each question in a separate session to ensure the independence of LLMs' answers.

\subsubsection{Experiment 2: Examine LLMs' Performance without Vis}
\label{subsec:novis+choices}
Given that the charts in the VLAT paper are derived from real datasets, it is imperative to investigate whether LLMs' responses towards visualization might be influenced by their existing knowledge from VLAT or other real-life datasets (\textbf{RQ1.2}). To this end, in this experiment, we investigated LLMs' question-answering performance without providing visualizations. In the pilot study where we followed the options structure in the VLAT~\cite{Lee:2017:VDV}, we found that for real-life questions, similarly, \gpt\ is likely to provide the answer \textit{Omit}. This is because \gpt\ claimed it \textit{``cannot provide real-time or location-specific information,''} making it difficult for us to analyze its responses.  To be consistent, we excluded the \textit{Omit} option and constrained LLMs to answer with the option letter only. Our prompt was structured as follows:
\textit{``You are a helpful assistant for answering questions. Please answer with the letter corresponding to the best option, or make a random guess if unsure. For instance, if option (a) is correct, please reply with (a).''}
We conducted 120 tests for each question in separate sessions similar to Section~\ref{subsec:vis+choices}, ensuring a counterbalanced order of all available choices. As there are 53 questions, we also ran 53 $\times$ 120 = 6,360 trials and shuffled the order with a different random seed than Experiment 1. We again recorded the time each LLM spent responding to each question. For \gpt, we opted to utilize the \texttt{gpt-4-turbo-preview} model as it has higher token limits and is comparable to the multimodal model \texttt{gpt-4-vision-preview} we used for analyzing visualizations. Similarly, we used the model \texttt{gemini-pro} to explore \gemini's knowledge when visualizations were absent.


\subsection{Analysis Methods}
We used the general public's performance as a baseline to evaluate LLMs' visualization literacy. We qualitatively compared human results to LLMs' results from Experiment 1. To address the \textbf{RQ1.1} about the effects of conditions, we developed the following three hypotheses:

\begin{itemize}[leftmargin=*]
\label{hypothesis}
    \item \textbf{H1}: \gpt\ would generally outperform \gemini.
    \item \textbf{H2}: LLMs perform better on the \textit{line chart}, \textit{bar chart}, \textit{pie chart}, \textit{scatterplot}, and \textit{area chart} compared to other visualizations.
    \item \textbf{H3}: LLMs perform better at \textit{Retrieve Value}, \textit{Find Extremum}, and \textit{Make Comparisons} than other tasks.
\end{itemize}

We formulated \textbf{H1} based on the fact that OpenAI was the first to market and had more time to collect user feedback from the public regarding \gpt. When forming \textbf{H2}, we believed that simple visualizations (\ie, visualizations that have minimal marks and channels) would be easier for LLMs to analyze. For \textbf{H3}, we assumed that these tasks were simpler than the other tasks within VLAT and, thus, would be easier for LLMs to accomplish.

For \textbf{RQ1.2}, we formulated another hypothesis based on the premise that if LLMs perform significantly better when visualizations are present, it suggests that LLMs answer questions with information presented in the visualization. This assumption is underpinned by the fact that all visualizations in our study contained random data values. We developed \textbf{H4} based on our intuition that having more information, such as the presence of the visualization, would allow LLMs to answer questions correctly more often, especially because the information in the visualization is necessary to answer the question correctly.

\begin{itemize}[leftmargin=*]
    \item \textbf{H4}: LLMs would perform better with visualizations.
\end{itemize}

Based on these hypotheses, we conducted hypothesis tests to examine the impact of four dimensions (visualization type, task type, LLM type, and visualization presence) on the LLMs' performances in answering questions. To achieve this, we modelled the results using logistic regression over these dimensions since our dependent variable (whether a question was answered correctly or not) is binary. Since all variables are categorical or binary, exploring them and their interactions was sufficient. The model is of the following form:

\vspace{-1.2em}
\begin{equation}
\label{eq:model}
    P(y=1|\textbf{x})=\text{logit}\left(\beta_0+\boldsymbol{\beta}_1^T\textbf{x}_1+\boldsymbol{\beta}_2^T\textbf{x}_2+\boldsymbol{\beta}_3^T\textbf{x}_3+\boldsymbol{\beta}_4^T\textbf{x}_4\right)
\end{equation}
\vspace{-1em}

\noindent where $y$ represents whether the question was answered correctly ($y=1$) or not ($y=0$), $\textbf{x}_k$ represents a vector of $k$-way interaction variables ($k \in \{1,2,3,4\}$ since we explored four dimensions), and $\boldsymbol{\beta}_k$ represents a vector of coefficients for each of these variables. For task types, we simplified the entries seen in \autoref{tab:results_ex1} by removing text in the parentheses.  For instance, \textit{Retrieve Value (Absolute Value)} became \textit{Retrieve Value}. After this simplification, a total of 49 unique visualization and task-type interactions remain, leading to a total of 629 variables (more details available in Appendix~\ref{app:varInterCalc}).

To find the best-performing model, we then conducted hyperparameter tuning by running multiple ten-fold cross-validations on the various penalties, solvers, and regularization values (see Appendix~\ref{sec:paramSelect} for details). After deciding the hyperparameters to use for the final model, we performed bootstrapping~\cite{tibshirani1993introduction} on the data to estimate the distributions of model coefficients and dependent variable probabilities. We resort to bootstrapping since it is a general method that can be used for hypothesis testing. Although traditional hypothesis methods like the difference-in-deviance method and Wald inference~\cite{montgomery2021introduction} can be used on the model coefficients, they cannot be applied to the dependent variable probabilities. We fitted the same model 1000 times, storing the coefficients for each bootstrapped model. These values are used to estimate the distributions, which in turn allow us to perform our hypothesis tests (more details in Appendix~\ref{app:bootstrapping}).

\begin{table*}[t]
    \centering
    \caption{Summary table of hypotheses, statistical methods, and results used to support/not support for \textbf{RQ1}.}
    \begin{tabular}{l|c|l|l|l}
        \hline
        RQ              &Hypothesis &Methods             &Results &Result Section(s)\\
        \hline
        \textbf{RQ1.1}  &\textbf{H1}
                        &Coefficient Analysis / Probability Difference Tests 
                        &Not supported 
                        &Sections~\ref{sec:RQ1_llm_human} and \ref{sec:coefResults}\\
        \textbf{RQ1.1}  &\textbf{H2}
                        &Coefficient Analysis
                        &Partially supported 
                        &Section~\ref{sec:vizTaskResults}                     \\
        \textbf{RQ1.1}  &\textbf{H3}
                        &Coefficient Analysis / Probability Difference Tests
                        &Partially supported 
                        &Section~\ref{sec:vizTaskResults}                     \\
        \textbf{RQ1.2}  &\textbf{H4}
                        &Probability Difference Tests
                        &Not supported
                        &Section~\ref{sec:vis_novis}\\
        \hline
    \end{tabular}
    \label{tab:hypotheses}
\end{table*}

For \textbf{H1}, we first examined the general performance of LLMs in terms of the coefficients and confidence intervals. To further examine the distinct performances of the two LLMs in completing various visualization tasks when visualizations were present, we conducted an additional hypothesis test. In particular, we took the difference of each pair of bootstrapped probabilities between \gpt\ and \gemini\ and performed a statistical test to see if their differences were equal to zero. The null hypothesis ($H_0$) was that the difference in probabilities was zero, while the alternative hypothesis ($H_1$) was that it was not zero. To be specific,
$H_0: \pi_{\text{GPT-4}}-\pi_{\text{Gemini}} = 0$; $H_1: \pi_{\text{GPT-4}}-\pi_{\text{Gemini}} \neq 0$. Here $\pi_i$ for $i \in \{\text{\gpt},\text{\gemini}\}$ represents the probability of LLM $i$ answering correctly. Initially, we tested whether $\pi_{\text{GPT-4}}$ and $\pi_{\text{Gemini}}$ were beta distributed. If they were, we planned to use the beta-difference distribution \cite{pham1993bayesian} to test if the difference between proportion pairs was greater than zero. If not, we intended to use the one-sided Wilcoxon signed-rank test on these differences with the estimated lower and upper bounds from the empirical cumulative distribution function (ECDF). This hypothesis was tested across all 49 visualization/task interactions.

We then tested if each coefficient was normally distributed for \textbf{H2} and \textbf{H3}. If it was, we performed a two-sided t-test to determine whether it was statistically different than zero. If not, we calculated a two-sided test using the Wilcoxon signed-rank test with estimated lower and upper bound values from the ECDF. Both methods used two-sided confidence intervals with $\alpha=0.05$. This allowed us to quantitatively address \textbf{RQ1.1} with the hypothesis test as $H_0: \beta_i = 0$; $H_1: \beta_i \neq 0$ where $\beta_i$ for $i \in \{1,\dots,629\}$ corresponds to the $i$th coefficient in the logistic model (see Appendix~\ref{app:bootstrapping}). A positive coefficient in the logistic regression model would indicate that a set of variables would more likely result in a correct answer, while a negative coefficient would indicate an incorrect answer.

For \textbf{H4}, we took the difference of each pair of bootstrapped probabilities for \gpt\ and \gemini\ separately to see how much the presence of the visualization affected the LLMs' ability. More explicitly, $H_0: \pi_{i,\text{Vis}}-\pi_{i,\text{No Vis}} = 0$; $H_1: \pi_{i,\text{Vis}}-\pi_{i,\text{No Vis}} > 0$ where $\pi_{i,j}$ for $i \in \{\text{\gpt},\text{\gemini}\}$ and $j \in \{\text{Vis},\text{No Vis}\}$ represents the probability of LLM $i$ answering correctly for visualization presence state $j$. Again, if $\pi_{i,\text{Vis}}$ and $\pi_{i,\text{No Vis}}$ were both beta distributed, we used the beta-difference distribution \cite{pham1993bayesian}. Otherwise, we used the one-sided Wilcoxon signed-rank test with estimated bounds from the ECDF. This hypothesis was tested across all 49 visualization/task interactions.

\subsection{Results}
In total, we collected 6,360 $\times$ 2 (models) $\times$ 2 (w/wo visualization) = 25,440 trials for Experiments 1 and 2, with each trial comprising the index, visualization type, task type, preset question, correct answer, LLMs' response, and experiment number. The overall hypothesis results are summarised in \autoref{tab:hypotheses}.

\begin{figure*}[h]
    \centering
    \begin{subfigure}{.494\textwidth}
        \includegraphics[width=\textwidth]{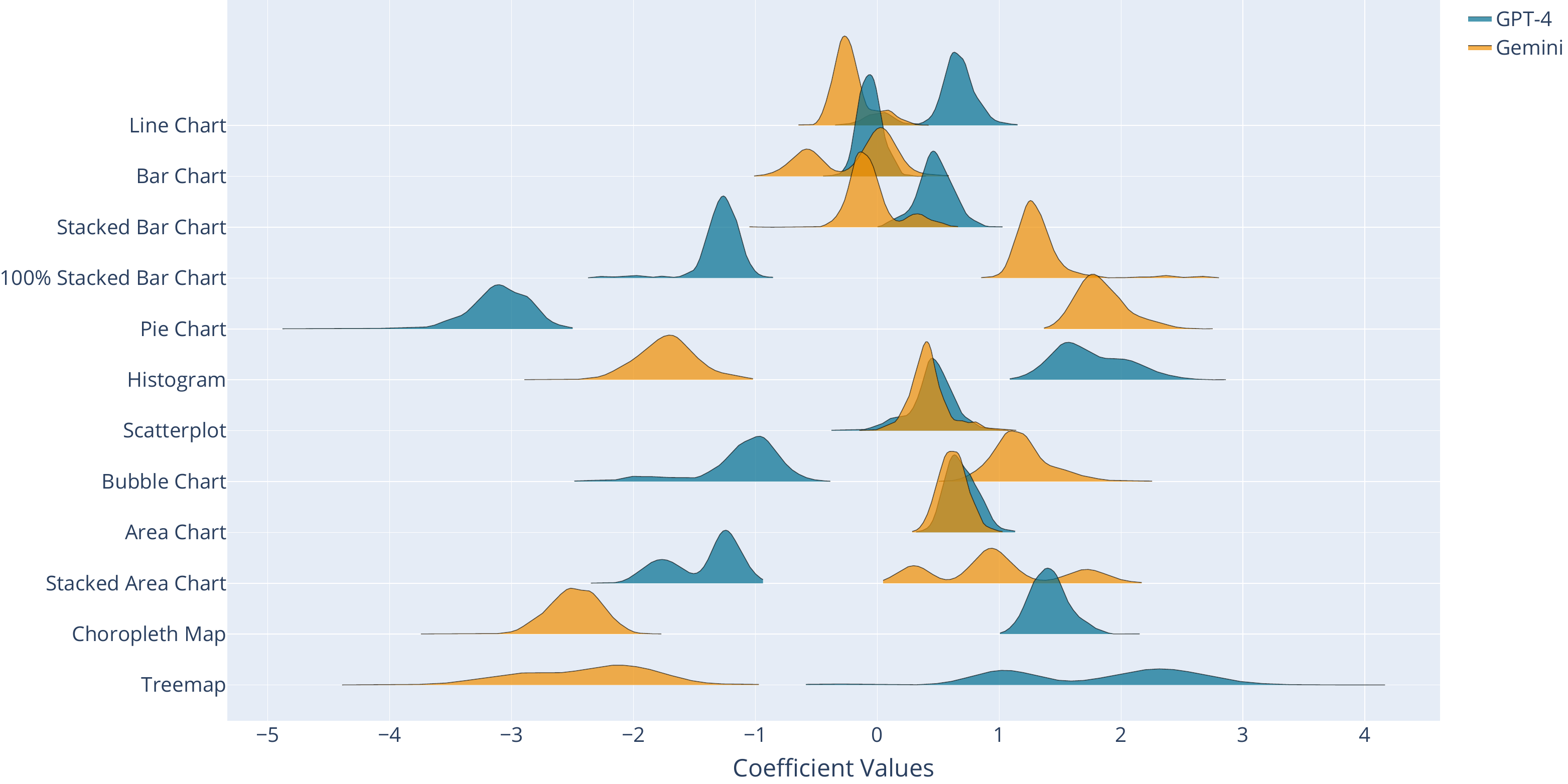}
        \caption{Visualization coefficients}
        \label{fig:chartCoef}
    \end{subfigure}
    \begin{subfigure}{.494\textwidth}
        \includegraphics[width=\textwidth]{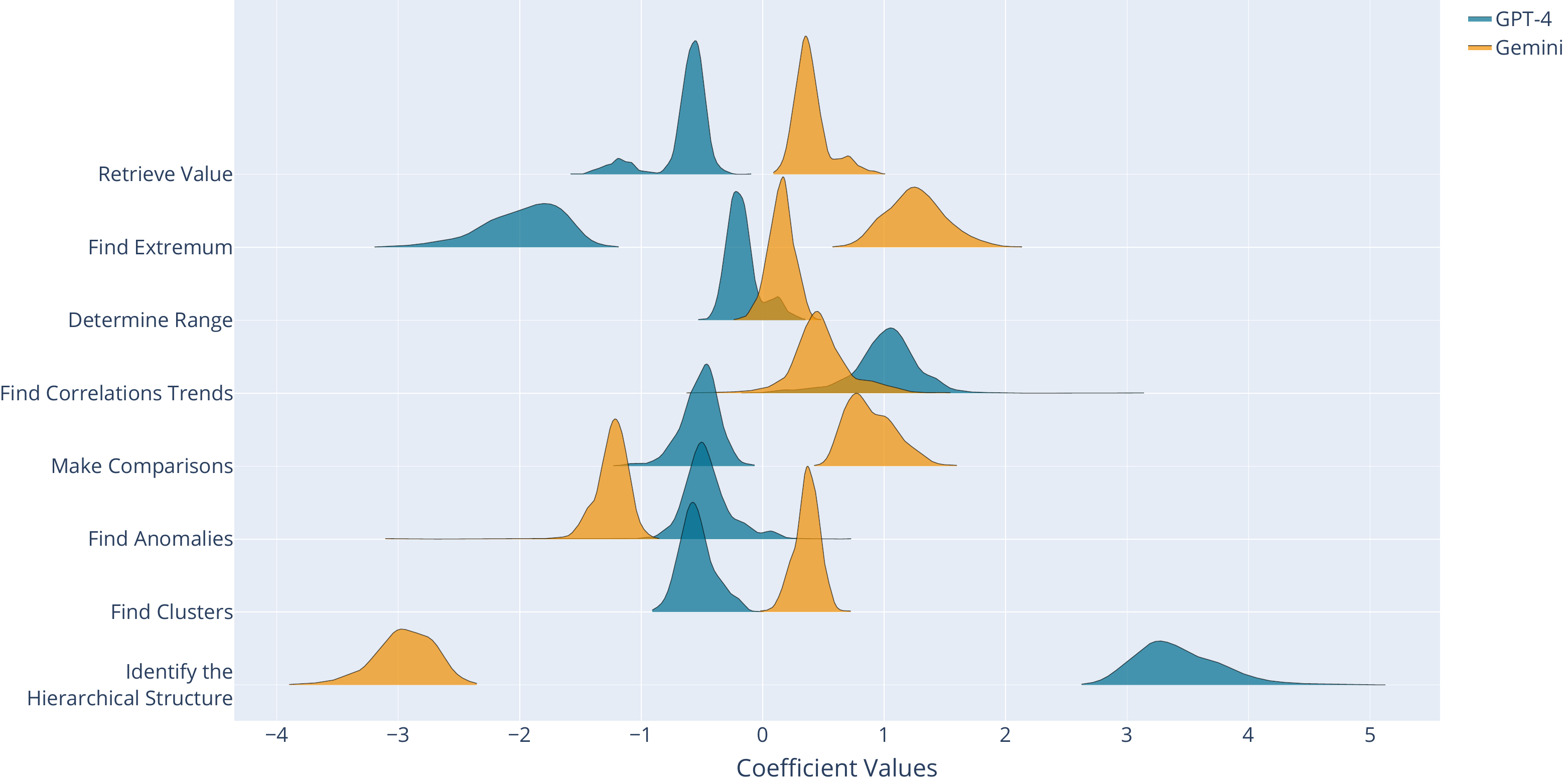}
        \caption{Task coefficients}
        \label{fig:taskCoef}
    \end{subfigure}
    \caption{Ridge plots of visualization and task bootstrapped coefficients when the visualization is present from the logistic regression. (a) illustrates coefficients of \gpt\ and \gemini\ for various visualization types. (b) shows the coefficients of these two LLMs for visualization tasks. A higher value indicates that the model performed better in this type of visualization (a) or visualization task (b).}
    \label{fig:chartTaskCoef}
\end{figure*}

\subsubsection{\textbf{RQ1} Visualization Literacy of LLMs vs. Humans}
\label{sec:RQ1_llm_human}
We compared the performance of LLMs with that of the general public, as we assume that if LLMs are proven to be more efficient, they could simulate human evaluators to provide initial assessments of visualizations. After compiling the human results from the original VLAT paper and our results from Experiments 1 and 2, we performed a qualitative comparison as presented in \autoref{tab:results_ex1}. Overall, \gpt\ and \gemini\ do not have visualization literacy levels comparable to humans. This means that if we use an LLM for initial evaluations, even the general public, and especially visualization experts, would need to adjust their approaches, as LLMs are likely to make more mistakes. Of the 53 visualization/task questions, \gpt\ outperformed humans on only 14 of these questions and \gemini\ on 15 questions. We also compared the LLMs' results with the accuracy of random chance, where all choices are evenly and randomly selected. \gpt\ performed better than random in 25 of the 53 questions, while \gemini\ exceeded 24 of them. The potential reason is that LLMs relied on their prior knowledge to answer questions, which may not align with the information presented in the visualizations.

\begin{table*}[]
    \caption{The comparisons between the accuracy rates of the general public from VLAT~\cite{Lee:2017:VDV}, the results of \gpt\ and \gemini\ in Experiment 1, and the expected accuracy rates when all choices are evenly and randomly chosen. For example, if there are four choices, the expected accuracy rate for random choices is 0.25. We color-encoded LLMs' results: \better{green} for much better than the performance of people (more than 0.05 higher), \close{yellow} for values that were close to people (higher or lower within 0.05), and \worse{red} for values much worse than those of people (more than 0.05 lower).}\vspace{-0.7em}
    \tabulinesep=1pt
    \begin{tabular}{>{\small}p{1.1cm}|>{\small}p{2.23cm}|>{\small}p{9.8cm}|>{\small}p{0.5cm}|>{\small}p{0.5cm}|>{\small}p{0.6cm}|>{\small}p{0.4cm}}
    \toprule    
    Visualization & Task & Stem & VLAT  & \gpt & \gemini & Random \\[-0.2em]
    \midrule
    \multirow{5}{*}{Line Chart} & Retrieve Value & What was the price of a barrel of oil in February 2015? & 0.95 & \worse{0.56} & \worse{0.25} & 0.25 \\[-0.2em]
                                & Find Extremum  & In which month was the price of a barrel of oil the lowest in 2015? & 0.97 & \worse{0.02} & \worse{0.22} & 0.25 \\[-0.2em]
                                & Determine Range & What was the price range of a barrel of oil in 2015? & 0.56 & \worse{0.23} & \worse{0.23} & 0.25 \\[-0.2em]
                                & Find Correlation/Trend & Over the course of the second half of 2015, the price of a barrel of oil was \underline{\hspace{1cm}}. & 0.98 & \worse{0.87} & \worse{0.42} & 0.33 \\[-0.2em]
                                & Make Comparisons & About how much did the price of a barrel of oil rise from June to September in 2015? & 0.77 & \better{0.87} & \worse{0.28} & 0.25 \\[-0.2em]
    \hline
    \multirow{4}{*}{Bar Chart} & Retrieve Value & What is the average internet speed in Japan? & 0.88 & \worse{0.40} & \worse{0.56} & 0.25 \\[-0.2em]
                               & Find Extremum & In which country is the average internet speed the fastest in Asia? & 0.98 & \worse{0.00} & \worse{0.01} & 0.25 \\[-0.2em]
                               & Determine Range & What is the range of the average internet speed in Asia? & 0.54 & \worse{0.29} & \better{0.68} & 0.25 \\[-0.2em]
                               & Make Comparisons & How many countries in Asia is the average Internet speed slower than South Korea? & 0.40 & \worse{0.20} & \worse{0.13} & 0.25 \\[-0.2em]
    \hline
    \multirow{5}{*}{\makecell[l]{Stacked \\ Bar Chart}} 
                                & \makecell[l]{Retrieve Value \\[-0.2em] (Absolute Value)} & What is the cost of peanuts in Las Vegas? & 0.38 & \worse{0.23} & \worse{0.25} & 0.25 \\[-0.2em]
                                & \makecell[l]{Retrieve Value \\[-0.2em] (Relative Value)} & About what is the ratio of the cost of a sandwich to the total cost of room service in Seattle? & 0.36 & \worse{0.04} & \worse{0.23} & 0.25 \\[-0.2em]
                                & Find Extremum & In which city is the cost of soda the highest? & 0.69 & \worse{0.17} & \worse{0.05} & 0.25 \\[-0.2em]
                                & \makecell[l]{Make Comparisons \\[-0.2em] (Absolute Value)} & The cost of water in Boston is higher than that of New York City. & 0.59 & \better{0.70} & \better{0.78} & 0.50 \\[-0.2em]
                                & \makecell[l]{Make Comparisons \\[-0.2em] (Relative Value)} & The ratio of the cost of peanuts to the cost of water in Las Vegas is higher than that of San Francisco. & 0.47 & \better{0.94} & \close{0.43} & 0.50 \\[-0.2em]
    \hline
    \multirow{3}{*}{\makecell[l]{100\% \\ Stacked \\ Bar Chart}} 
                                & \makecell[l]{Retrieve Value \\[-0.2em] (Absolute Value)} & What is the approval rating of Republicans among the people who have the education level of Postgraduate Study? & 0.49 & \worse{0.00} & \better{0.79} & 0.25 \\[-0.2em]
                                & \makecell[l]{Find Extremum \\[-0.2em] (Relative Value)} & What is the education level of people in which the Democrats have the lowest approval rating? & 0.90 & \worse{0.00} & \worse{0.21} & 0.25 \\[-0.2em]
                                & \makecell[l]{Make Comparisons \\[-0.2em] (Relative Value)} & The approval rating of Republicans for the people who have the education level of Some College Degree is lower than that for the people who have the education level of Postgraduate Study. & 0.54 & \worse{0.15} & \better{0.98} & 0.50 \\[-0.2em]
    \hline
    \multirow{3}{*}{Pie Chart} 
                            & \makecell[l]{Retrieve Value \\[-0.2em] (Relative Value)} & About what is the global smartphone market share of Huawei? & 0.72 & \worse{0.00} & \worse{0.36} & 0.25 \\[-0.2em]
                            & \makecell[l]{Find Extremum \\[-0.2em] (Relative Value)} & In which company is the global smartphone market share the smallest? & 0.98 & \worse{0.00} & \worse{0.30} & 0.25 \\[-0.2em]
                            & \makecell[l]{Make Comparisons \\[-0.2em] (Relative Value)} & The global smartphone market share of Lenovo is larger than that of Samsung. & 1.00 & \worse{0.00} & \worse{0.43} & 0.50 \\[-0.2em]
    \hline
    \multirow{3}{*}{Histogram}
                            & \makecell[l]{Retrieve Value \\[-0.2em] (Derived Value)} & How many people have rated the taxi between 4.0 and 4.2? & 0.84 & \worse{0.38} & \worse{0.18} & 0.25 \\[-0.2em]
                            & \makecell[l]{Find Extremum \\[-0.2em] (Derived Value)} & What is the rating that the people have rated the taxi the most? & 0.94 & \worse{0.06} & \worse{0.13} & 0.25 \\[-0.2em]
                            & \makecell[l]{Make Comparisons \\[-0.2em] (Derived Value)} & More people have rated the taxi between 4.6 and 4.8 than between 4.2 and 4.4. & 0.86 & \close{0.89} & \worse{0.00} & 0.50 \\[-0.2em]
    \hline
    \multirow{7}{*}{Scatterplot}
                            & Retrieve Value & What is the weight for the person who is 165.1 cm tall? & 0.85 & \worse{0.20} & \worse{0.49} & 0.25 \\[-0.2em]
                            & Find Extremum & What is the height for the tallest person among the 85 males? & 0.76 & \better{0.85} & \better{1.00} & 0.25 \\[-0.2em]
                            & Determine Range & What is the range in weight for the 85 males? & 0.53 & \better{0.86} & \better{0.76} & 0.25 \\[-0.2em]
                            & Find Anomalies & What is the height for a person who lies outside the others the most? & 0.42 & \worse{0.00} & \worse{0.00} & 0.25 \\[-0.2em]
                            & Find Clusters & A group of males are gathered around the height of 176 cm and the weight of 70 kg. & 0.90 & \worse{0.06} & \worse{0.50} & 0.50 \\[-0.2em]
                            & Find Correlation/Trend & There is a negative linear relationship between the height and the weight of the 85 males. & 0.52 & \better{1.00} & \worse{0.00} & 0.50 \\[-0.2em]
                            & Make Comparisons & The weights for males with the height of 188 cm are all the same. & 0.79 & \better{1.00} & \better{1.00} & 0.50 \\[-0.2em]
    \hline
    \multirow{4}{*}{Area Chart}
                            & Retrieve Value & What was the average price of a pound of coffee beans in June 2013? & 0.75 & \worse{0.28} & \worse{0.13} & 0.25 \\[-0.3em]
                            & Find Extremum & When was the average price of a pound of coffee beans at minimum? & 0.44 & \better{0.57} & \better{0.73} & 0.25 \\[-0.3em]
                            & Determine Range & What was the range of the average price of a pound of coffee beans between January 2013 and December 2014? & 0.38 & \worse{0.23} & \worse{0.13} & 0.25 \\[-0.3em]
                            & Find Correlation/Trend & Over the course of 2013, the average price of a pound of coffee beans was \underline{\hspace{1cm}}. & 0.94 & \better{1.00} & \better{1.00} & 0.33 \\[-0.2em]
    \hline
    \multirow{6}{*}{\makecell[l]{Stacked \\ Area Chart}}
                            & \makecell[l]{Retrieve Value \\[-0.2em] (Absolute Value)} & What was the number of girls named `Amelia' in 2010 in the UK? & 0.15 & \better{0.36} & \better{0.35} & 0.25 \\[-0.2em]
                            & \makecell[l]{Retrieve Value \\[-0.2em] (Relative Value)} & About what was the ratio of the number of girls named `Amelia' to those named `Isla' in 2014 in the UK? & 0.25 & \worse{0.17} & \close{0.30} & 0.25 \\[-0.2em]
                            & Find Extremum & Over the course of years between 2009 and 2014, when was the number of girls named `Amelia' at the maximum? & 0.97 & \worse{0.00} & \worse{0.01} & 0.25 \\[-0.2em]
                            & Find Correlation/Trend & The number of girls named `Isla' was \underline{\hspace{1cm}} from 2009 to 2012. & 0.96 & \worse{0.00} & \worse{0.72} & 0.33 \\[-0.2em]
                            & \makecell[l]{Make Comparisons \\[-0.2em] (Absolute Value)} & In the UK, the number of girls named `Amelia' in 2014 was more than it was in 2013. & 0.20 & \better{0.51} & \better{0.99} & 0.50 \\[-0.2em]
                            & \makecell[l]{Make Comparisons \\[-0.2em] (Relative Value)} & Over the course of years between 2009 and 2014, the number of girls named `Isla' was always more than `Olivia'. & 0.24 & \better{1.00} & \better{1.00} & 0.50 \\[-0.2em]
    \hline                            
    \multirow{7}{*}{\makecell[l]{Bubble \\ Chart}}
                            & Retrieve Value & What is the total length of the metro system in Beijing? & 0.41 & \worse{0.02} & \worse{0.15} & 0.25 \\[-0.2em]
                            & Find Extremum & Which city’s metro system has the largest number of stations? & 0.69 & \worse{0.00} & \worse{0.11} & 0.25 \\[-0.2em]
                            & Determine Range & What is the range of the total length of the metro systems? & 0.29 & \close{0.25} & \close{0.26} & 0.25\\[-0.2em]
                            & Find Anomalies & Which city’s metro system does lie outside the relationship between the total system length and the number of stations most? & 0.53 & \close{0.53} & \worse{0.06} & 0.25 \\[-0.2em]
                            & Find Clusters & A group of the metro systems of the world has approximately 200 stations and around a 200 km system length. & 0.59 & \worse{0.38} & \worse{0.48} & 0.50 \\[-0.2em]
                            & Find Correlation/Trend & In general, the ridership of the metro system increases as the number of stations increases. & 0.26 & \better{0.93} & \better{1.00} & 0.50 \\[-0.2em]
                            & Make Comparisons & The metro system in Paris has more ridership than the metro system in New York City. & 0.80 & \worse{0.00} & \worse{0.09} & 0.50 \\[-0.2em]
    \hline
    \multirow{3}{*}{\makecell[l]{Choropleth \\ Map}}
                            & \makecell[l]{Retrieve Value \\[-0.2em] (Approximate Value)} & What was the unemployment rate for Indiana (IN) in 2015? & 0.24 & \worse{0.00} & \worse{0.00} & 0.25 \\[-0.2em]
                            & \makecell[l]{Find Extremum \\[-0.2em] (Approximate Value)} & In which state was the unemployment rate the highest in 2015? & 0.97 & \close{0.93} & \worse{0.72} & 0.25 \\[-0.2em]
                            & \makecell[l]{Make Comparison \\[-0.2em] (Approximate Value)} & In 2015, the unemployment rate for Arizona (AZ) was higher than that of Oklahoma (OK). & 0.92 & \worse{0.80} & \worse{0.00} & 0.50 \\[-0.2em]
    \hline
    \multirow{3}{*}{Treemap}
                            & \makecell[l]{Find Extremum \\[-0.2em] (Relative Value)} & For which website was the number of unique visitors the largest in 2010? & 0.68 & \worse{0.01} & \worse{0.00} & 0.25 \\[-0.2em]
                            & \makecell[l]{Make Comparison \\[-0.2em] (Relative Value)} & The number of unique visitors for Target was more than that of Ask in 2010. & 0.42 & \worse{0.03} & \better{0.53} & 0.50 \\[-0.2em]
                            & \makecell[l]{Identify the \\[-0.2em] Hierarchical Structure} & Amazon is nested in the Computer category. & 0.92 & \better{1.00} & \worse{0.00} & 0.50 \\
    \hline
    \end{tabular}
    \label{tab:results_ex1}
\end{table*}

\subsubsection{\textbf{H1} Visualization Literacy of \gpt\ and \gemini}
\noindent\textbf{Coefficient Results}
\label{sec:coefResults}
Out of the 629 variables and interactions, 623 were found to be statistically significant, indicating that the majority of these combinations have an effect on the LLMs' performance. We reported the detailed coefficients in Appendix~\ref{app:coefficients}. When focusing on the presence of visualizations, both LLMs jointly perform poorly ($\bar{\beta} \approx -0.5051$, where $\bar{\beta}$ represents the mean of the bootstrapped coefficients). An individual examination of each LLM individually revealed that \gpt\ performed relatively well ($\bar{\beta} \approx 0.1325$) while \gemini\ performed poorly ($\bar{\beta} \approx -0.6376$).

\noindent\textbf{Bootstrapped Probabilities Results}
When we directly compared the LLMs by examining the difference between their bootstrapped probabilities, we found that of the 49 visualization/task interactions, 42 were statistically significant. Among these, 17 showed that \gpt\ performed better than \gemini, while 25 indicated that \gemini\ performed better than \gpt. The remaining 7 show no difference (see Appendix~\ref{app:bootstrapLLM}). This is surprising as the general coefficient analysis we reported in the previous paragraph seemed to indicate that \gpt\ outperforms \gemini\ when the visualization is present. 

Although some of these results seem to conflict with the previous section's conclusion that \gpt\ performs better at visualization literacy questions than \gemini\ (\ie, \gemini\ had more positive tasks and interaction coefficients than \gpt), we believe this is a classic example of Simpson's paradox \cite{simpson1951interpretation}, where the subsets show an opposing trend compared to the full dataset.  This is especially true for our model since the coefficients in regression models with binary variables will sum to the more general coefficient. The ridge plots suggest that the largest contributing factor as to why \gpt\ performed better than \gemini\ overall seems to be attributed to the large difference in performance for the task \textit{Identify the Hierarchical Structure}.  Without this task, it appears that \gemini\ would be better than \gpt\ overall.

\subsubsection{\textbf{H2 \& H3} LLMs in Varied Visualizations and Tasks}\label{sec:vizTaskResults}
All of the insignificant interaction coefficients were cases where the visualization was absent except for one, which means that these conditions do not contribute to the LLMs' responses. Further, all of these coefficients were interaction coefficients between visualization and task types, meaning that all of the single interaction coefficients for visualization and task types were statistically significant. We report the statistically insignificant coefficients in Appendix~\ref{app:coefficients}. We visualized the detailed performances of LLMs in diverse conditions in \autoref{fig:chartTaskCoef}.

Starting with visualization type, when the visualization was present, 7 out of the 12 coefficients for both LLMs (\ie, no interaction with tasks or models) were positive. While examining which coefficients were positive, we found that the coefficients for the \textit{line chart}, \textit{scatterplot}, and \textit{area chart} seemed to support \textbf{H2}, but surprisingly, the \textit{pie chart} was not included (see \autoref{fig:chartCoef}). Complicated visualizations like the \textit{stacked bar chart}, \textit{100\% stacked bar chart}, \textit{histogram}, and \textit{bubble chart} had significant positive coefficients, contrary to our expectations, as we believed LLMs would have had a hard time parsing their marks and channels. For the \gpt\ interaction coefficients, 7 were positive, while for \gemini, 6 were positive.

For task types, 3 out of the 8 task coefficients were positive for both LLMs. We found that only the positive coefficient that supported \textbf{H3} was \textit{Make Comparisons}. The other tasks with positive coefficients (\textit{Find Correlation/Trends} and \textit{Identify the Hierarchical Structure}) seemed more difficult, which surprised us. For \gpt, only 2 coefficients were positive, while \gemini\ had 6 positive coefficients. \edit{This suggests that for contouring tasks such as \textit{Retrieve Value}, \textit{Find Extremum}, \textit{Make Comparisons}, and \textit{Find Clusters}, \gemini\ was likely to outperform \gpt\ and could be the preferred choice.} We visualized the coefficients' distributions in the ridge plot in \autoref{fig:taskCoef}.

\begin{figure}
    \centering
    \includegraphics[width=\columnwidth]{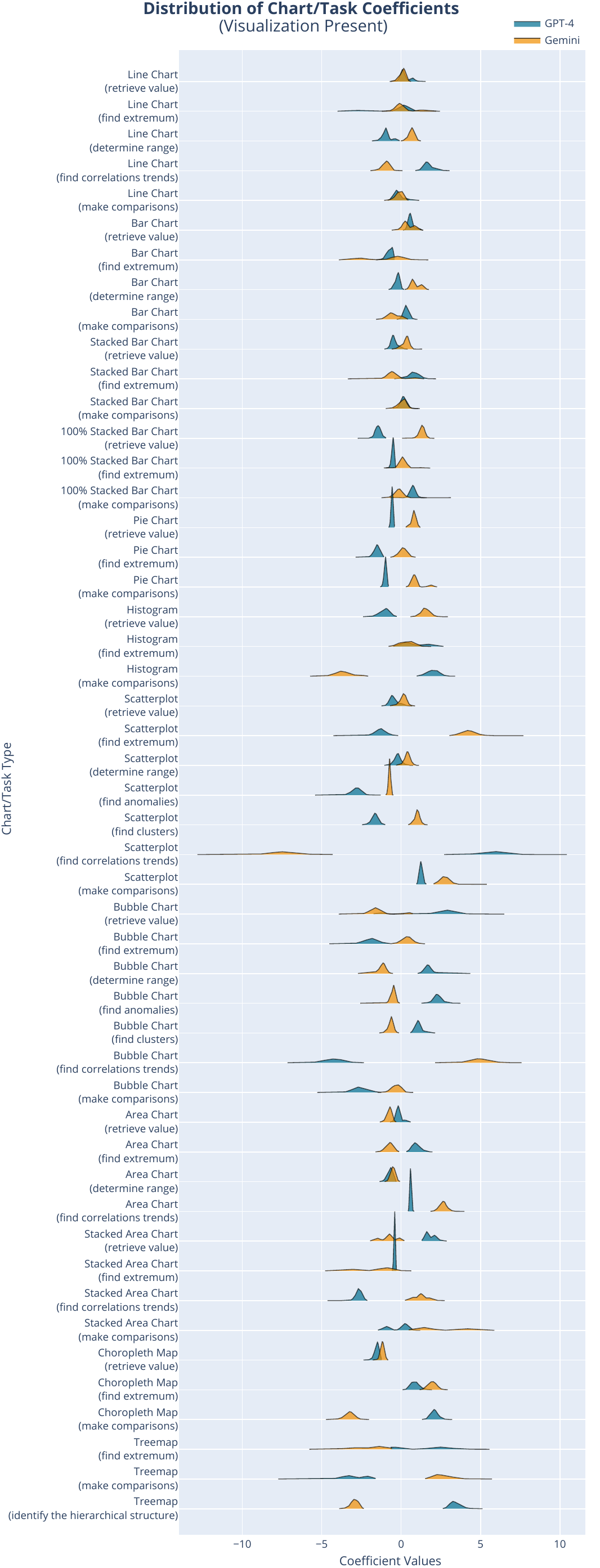}
    \caption{Ridge plots of the bootstrapped visualization/task interaction coefficients when the visualization is present from the logistic regression for \gpt\ and \gemini.}
    \label{fig:chartTaskInterCoef}
\end{figure}

Finally, for visualization/task interactions, 25 out of the 49 interaction coefficients were positive for both LLMs. For \gpt, 22 interaction coefficients were positive, while for \gemini, 26 were positive (though one was statistically insignificant). To illustrate these model-specific interactions in depth, we visualized the detailed performance in a ridge plot of the interaction coefficients (see \autoref{fig:chartTaskInterCoef}). This visualization offers insights into each model's performance, helping to guide selection based on task requirements. For example, when users focused on \textit{Finding Correlations Trend} in \textit{scatterplot} \edit{or \textit{Making Comparisons} in \textit{choropleth map}}, \gpt\ showed stronger alignment, suggesting it might be the better choice for such tasks. \edit{Conversely, when experts are interested in \textit{Finding Extremum} in \textit{scatterplot} or \textit{Finding Correlations Trend} in \textit{bubble chart}, it might be more effective to use \gemini.} This approach allows people to select models based on specific scenarios, optimizing their use of LLMs for enhanced task accuracy.

\subsubsection{\textbf{H4} LLMs' Performances with Vis vs. No Vis}
\label{sec:vis_novis}
When we compared the results on visualization presence by taking the difference between their bootstrapped probabilities, 18 of the 49 visualization/task interactions were statistically significant for \gpt, and 20 were statistically significant for \gemini\ (more details in Appendix~\ref{app:visdiffResults}). Since the majority of visualization/task interactions showed no significance for both \gpt\ and \gemini, we were unable to support \textbf{H4}, suggesting that LLMs mostly rely on their knowledge base to answer visualization questions despite only needing the information in visualizations to answer the question.  Although our results show which visualization/task questions each LLM can answer regardless of the context of the question, this does not fully address how to prevent LLMs from using their knowledge base when answering questions concerning data visualizations.  

\section{Further Analysis of LLMs in Visualization Understanding}
We reflected on the results from Section~\ref{sec:assessingVL} and attempted to further explore the limitations of using LLMs to interpret visualization in the future. We envision two scenarios. The first one is about visualization evaluation. A designer creates a visualization and seeks an initial evaluation of its clarity and comprehensiveness. To mitigate the potentially high costs of recruiting human evaluators, the designer considers using LLMs for this task. In this context, the designer should decide whether to provide any constraints, such as answer choices, to LLMs. The other scenario is about visualization reading. An individual encounters a visualization without prior knowledge of its content and requests insights from the LLMs. Here, it is unlikely that answer choices would be provided, making LLMs' performance in open-ended questions crucial.
The existing visualization literacy assessments, such as VLAT~\cite{Lee:2017:VDV} and MiniVLAT~\cite{Pandey:2023:MiniVLAT}, streamline the answer choices to expedite the evaluation process and enhance practicality and time-effectiveness. However, as LLMs' responses do not entail human effort, it is practical to use LLMs to answer open-ended questions related to visualizations.

To this end, we conducted two experiments to assess LLMs' performances when no answer choices were provided. As detailed in Section~\ref{sec:vis_novis}, our findings revealed that LLMs heavily relied on their pre-existing knowledge to answer questions, especially in some specific topics, \eg, politics. To delve deeper into this aspect, we introduced a follow-up experiment focusing on decontextualized visualizations. In summary, we were curious about the following sub-research questions when using LLMs in visualization interpretation: 
\begin{itemize}
    \item \textbf{RQ2.1}: Are LLMs limited by answer choices?
    \item \textbf{RQ2.2}: Are LLMs limited by contextualized visualizations?
\end{itemize}
As a whole, we aimed to derive potential strategies for answer choices and contexts to enhance LLMs' effectiveness in reading and understanding visualizations.

\subsection{Experiments}
To fully address \textbf{RQ2.1}, we designed Experiments 3 and 4, using the same visualizations and questions outlined in Section~\ref{subsec:vis_design}. Our goal was to determine if the absence of answer choices would impact LLMs' performance. As for \textbf{RQ2.2}, we conducted Experiment 5 with the contextualized visualizations. We included the answer choices for better comparison with the results from Experiment 1.

\vspace{-0.3em}
\subsubsection{Experiment 3: Examine LLMs' Choice-Free Performance}
\label{subsec:vis+nochoices}
In this experiment, we asked \gpt\ and \gemini\ the same questions as Experiment 1 but this time without providing answer options. The goal was to determine whether the absence of predefined choices would affect the models' performance across different types of tasks. To ensure a focused analysis, we instructed the models to respond as succinctly as possible. However, even when instructed to provide short responses, \gpt\ can still generate sentences, which can be overwhelming, especially considering the thousands of trials we would collect in this experiment. After several attempts, we used the prompt as follows:
\textit{``You are a helpful assistant for analyzing data visualizations. Please answer with the best response in one word.''}
In this experiment, we also used the \gpt\ model \texttt{gpt-4-vision-preview} and \gemini\ model \texttt{gemini-pro-vision} and conducted 120 tests for all 53 questions in separate sessions with a random order to maintain consistency. While we did not have answer choices to counterbalance, we recorded the time taken for each question for further comparisons.

\vspace{-0.3em}
\subsubsection{Experiment 4: Examine LLMs' Performance without Visualization and Choices}
\label{subsec:novis+nochoices}
For consistency, in this experiment, we tested LLMs without providing answer choices or visualizations to determine if LLMs' responses would differ when visualizations were not present compared to the results from Experiment 3. Without visualizations, we expected the LLMs to struggle with answering the questions, as they would lack the necessary data to form accurate responses. We used the following prompt: 
\textit{``You are a helpful assistant for answering questions. Please answer with the best response in one word.''} 
We also repeated 120 times for each question, shuffled the question order, and recorded the time each question took. We used the \texttt{gpt-4-turbo-preview} and \texttt{gemini-pro} models.

\vspace{-0.3em}
\subsubsection{Experiment 5: Examine LLMs' Performance of Decontextualized Visualization}
\label{sec:decontext}
To investigate the impact of visualization context, we conducted a follow-up study where we removed all proper nouns in the visualizations. We replaced all the city names with ``City N'', product names such as ``Oil'' and ``Peanuts'' with ``Product N'', country names with ``Country N'', and political party names with ``Party A''. We excluded three visualizations from this experiment: \textit{histogram}, \textit{scatterplot}, and \textit{choropleth map}. The \textit{histogram} and \textit{scatterplot} had previously used random data unrelated to specific properties (a small group of taxi ratings and 85 individuals' heights and weights), while the \textit{choropleth map} contained inherent geographic information that could not be decontextualized. Consequently, we had a total of 40 questions. We removed the \textit{Omit} option for consistency and better comparisons with previous experiments. We tested both \gpt\ and \gemini\ in answering these questions with choices, as we did in Experiments 1 and 2, but with decontextualized visualizations. In total, we had 40 (questions) $\times$ 120 (times) $\times$ 2 (models) $\times$ 2 (w/wo visualizations) = 19,200 trials.

\subsection{Analysis Methods}
As Experiments 3 and 4 dealt with no-choice questions for \textbf{RQ2.1} while Experiment 5 involved questions with choices for \textbf{RQ2.2}, we analyzed their results separately.

\subsubsection{Analysis for Experiments 3 and 4}
\label{subsubsection:analysis_for_experiments_3_and_4}
Out of the 53 questions, 30 required textual answers initially (including two questions asking about a specific year or month), while the remaining 23 sought numerical responses. Ideally, there would be 14,400 textual trails and 11,040 numerical trails for \gpt\ and \gemini. However, large language models (LLMs) can generate a wide range of responses, \eg, giving textual responses to questions requiring numerical answers. As such, we conducted separate analyses for questions demanding numerical and textual responses. To facilitate categorization and analysis, we standardized similar terms. For example, we mapped ``Inverse'' to ``False,'' ``Correct'' to ``True,'' and ``Similar'' to ``1/1.'' Additionally, we made judgments to some answers, such as mapping ``likely'' to ``True'' and changing ``Doctoral study'' to ``Postgraduate Study'' to align with our preset choices. We also created an ``Error'' category named to include all unexpected responses. This category includes ``Random'' responses, which were generated with random characters or nonsensical content; ``Vague'' responses, which were vague and difficult to interpret (\eg, answering `fast' to a question about the average internet speed); ``Unknown'' responses, where LLMs indicated their inability to provide an answer due to uncertainty or lack of evidence (\eg, ``unsure'', ``indeterminate'', and empty responses) or LLMs repeated the word from the question (\eg, responding `height' for a question about a person's height); and ``Prompt Engineering'' responses, which likely occurred due to the unsuitable questions as prompts. All these unexpected replies were considered wrong when evaluating the accuracy of LLMs in answering open questions. In addition, if a response provided multiple answers, even though it covered the right answer, we marked it wrong. It is worth noting that as we generated random values in creating visualizations, the correct answers refer to the values that align with our generated data rather than with real-world data.

\vspace{0.1em}
\noindent\textbf{Textual Responses Analysis}
For the questions with textual answers, we collected all the responses, transformed them into standardized terms, and compared them to the correct answers, assigning a score of 1 for matched responses and 0 for incorrect ones.

\vspace{0.1em}
\noindent\textbf{Numerical Response Analysis}
In all the questions requesting numerical answers, there are two categories: those asking for single values and those asking for value ranges. The latter category, in particular, was designed for the task \textit{Determine Range}.  

\noindent\emph{Two Special Cases Analysis.}
\label{para:two_special_cases}
During our analysis of numerical answers, we identified two questions where the correct answers were specified as ranges due to the categorical nature of the data. Rather than \textit{Determining Range}, these two questions were designed for other types of visualization tasks. We evaluated whether the responses fell within the correct answer ranges of these questions. If they did, we marked them as correct; otherwise, we marked them as incorrect.

\noindent\emph{Single Numerical Value Analysis.}
For the questions demanding a single value, we calculated the relative error between the response and the correct answer (\autoref{eq:relative_error}) to determine how the responses differ.
\begin{equation}
\label{eq:relative_error}
    \text{Relative Error} = \left|\frac{\text{Response - Correct Answer}}{\text{Correct Answer}}\right|
\end{equation}
The smaller the relative error is, the more accurate the LLMs are. 

\vspace{-0.2em}
\noindent\emph{Numerical Range Analysis.}
For the range-determining questions (5 questions), we applied multiple metrics to gain a comprehensive picture of the overlap between the correct range ($A$) and the LLM-predicted range ($B$). The four metrics we calculated to assess the overlap are: Percentage Overlap, Jaccard Index, Sørensen-Dice Coefficient, and Overlap Coefficient. Due to the mathematical properties of these functions, they are commutative, meaning the order of $A$ and $B$ does not affect the outcome. Percentage Overlap quantifies the direct proportion of overlap, emphasizing the actual shared range and encouraging precise alignment with the correct range. Jaccard Index offers a ratio of intersection to union, penalizing answers that extend far beyond the correct range, thus balancing precision and inclusivity~\cite{Costa:2021:Jaccard}. Sørensen-Dice Coefficient doubles the importance of the intersection, promoting answers that closely match the correct range while still considering the total size of both ranges~\cite{Jackson:1989:Similarity}. Overlap Coefficient focuses on the ratio of overlap to the smaller of the two ranges, highlighting the importance of covering the correct range without unnecessary extension~\cite{Vijaymeena:2016:Survey}. We reported the detailed equations of the four metrics in Appendix~\ref{app:numerical_range_ana}.


\subsubsection{Analysis for Experiment 5}
The purpose of Experiment 5 is to investigate whether using decontextualized visualizations could force LLMs to rely more on the information in the visualization. We aimed to compare the results with those from Experiments 1 and 2 qualitatively. To this end, we calculated the accuracy rates in all conditions.

\subsection{Results}

\subsubsection{\textbf{RQ2.1} Results of Experiments 3 and 4}
Among all the 6,360 $\times$ 2 (models) $\times$ 2 (w/wo visualization) = 25,440 responses, we identified 348 responses (about 1.37\%) as ``Random.'' A total of 2,888 responses were classified as ``Vague'', accounting for 11.35\% of the total responses. There are 1,268 ``Unknown'' replies, representing 4.98\% of the dataset. Also, 875 trials fall into ``Prompt Engineering''. Therefore, about 21.14\% responses are ``Error'' for both numerical and textual responses. Given our separate analysis of textual and numerical responses, we reported the results separately below.

\vspace{0.1em}
\noindent\textbf{Textual Responses}
Out of 14,119 valid responses that were textual instead of numerical, 1,290 responses fell into the category ``Error,'' and 4,587 were correct, resulting in an accuracy rate of 32.49\%. We further analyzed the performances of \gpt\ and \gemini\ separately. When visualizations were provided, the accuracy rates of both models significantly increased. Without visualization, \gpt\ performed worse than \gemini, with an accuracy rate of 28.33\% compared to Gemini's is 29.73\%. However, with visualization, \gpt\ outperformed \gemini, achieving accuracy rates of 38.27\% and 32.39\%, respectively.

\vspace{0.1em}
\noindent\textbf{Numerical Response}
\noindent\emph{Two Special Cases.}
As we mentioned in Section~\ref{para:two_special_cases}, there are two special questions where the answers are present in ranges. These two questions were ``What was the unemployment rate for Indiana (IN) in 2015?'' (answer: 1.1\%-2.3\%) and ``What is the rating that the people have rated the taxi the most?'' (answer: 4.0-4.2.) 
For the first question, out of 480 tests (with or without visualization), \gemini\ provided a correct answer only once, while all responses from \gpt\ were wrong. This was likely due to their background knowledge, as they both provided answers that were close to reality. For the second question, without visualization, both LLMs performed poorly, with correct answers given only twice. However, when visualizations were provided, \gpt\ correctly provided a number within the range for all 120 tests, while \gemini\ did so for 119 tests.

\begin{figure}[h]
    \centering
    \includegraphics[width=0.45\textwidth]{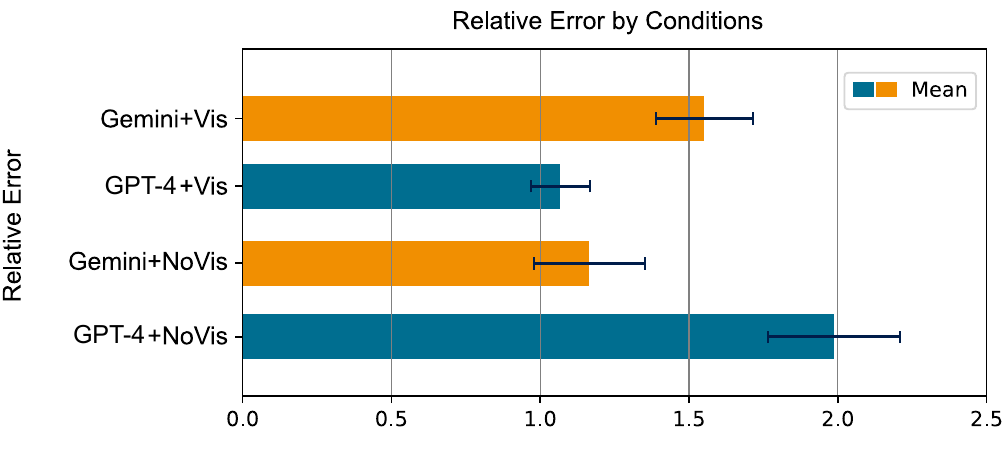}\vspace{-1ex}
    \caption{The 95\% confidence interval of relative errors for single numerical value responses across all conditions.}\vspace{-.5ex}
    \label{fig:relativeError}
    \vspace{-4mm}
\end{figure}

\noindent\emph{Single Numerical Value Results.}
We received 4,614 valid responses for questions demanding a single value. The total average relative error for the single-value questions is 1.44. We visualized the detailed results in \autoref{fig:relativeError}, which shows \gpt\ outperforming Gemini, especially when visualizations were provided. Interestingly, \gpt's performance improved significantly with visualizations, while \gemini's performance worsened under the same conditions. This suggests that \gemini\ is more likely to rely on its pre-existing knowledge for tasks involving the retrieval of single numerical values.

\noindent\emph{Numerical Range Results.}
In general, including a visualization tended to simplify the task for LLMs, leading to improved performance metrics compared to when visualizations were absent. \gpt\ generally excelled across all visualization types except for the \textit{area chart}, where \gemini\ attained a median overlap percentage of approximately 70\% when the \textit{area chart} was present, contrasting with a median of 0\% in all other conditions. \gemini\ yielded no valid responses when no visualization was present, resulting in a median of 0\%. For the \textit{bubble chart}, neither LLM performed consistently well under any metric, although occasional outliers demonstrated some level of proficiency. \gpt\ showed strong performance in the \textit{scatterplot}, whereas other combinations of LLM and visualization presence resulted in near-zero median performances. In the \textit{line chart}, \gpt\ demonstrated significant competency, with medians exceeding 80\% across all metrics under the condition of without visualization, which approached nearly 100\% with the addition of visualizations. Conversely, results for \gemini\ varied widely in the condition where visualization was present, with metrics spanning the full range from 0 to 100\%. For the \textit{area chart}, all conditions, except \gemini\ when visualizations were present, resulted in nearly zero values. Lastly, in the \textit{bar chart} evaluations, the only combination that exhibited some level of reliability was \gpt, which dealt with visualization, with all other conditions showing medians close to zero. We summarized the detailed results in Appendix~\ref{app:numerical_range_ana}. 

Therefore, providing choices may enhance the accuracy rates of LLMs in answering visualization-related questions in general, as there are too many unexpected responses. However, it also entails additional costs in designing these choices. In addressing \textbf{RQ2.1}, it seems that providing answer choices does not constrain LLMs, but instead, it guides them to provide answers that align with our expectations.

\vspace{-0.2em}
\subsubsection{\textbf{RQ2.2} Accuracy Rates in Experiment 5}
The average accuracy rate of \gpt\ in reading decontextualized visualizations is 41.96\%, compared to 30.90\% for the eight contextualized visualizations in Experiment 1. Among all the visualizations, the \textit{100\% stacked bar chart}, where party names were hidden, showed the most significant improvement in performance (from the average accurate percentage of 4.17\% to 36.81\%). Due to the limited number of task types, \eg, only one instance of \textit{Find Anomalies}, we cannot draw statistical conclusions from the results. Therefore, we focused on examining the absolute difference when grouping the task types. The most significant different task was observed in \textit{Find Extremum} task for \gpt. As for Gemini, the average accuracy rate when dealing with contextualized visualizations is 38.98\%, while in the follow-up study with decontextualized visualization, it reaches an average accuracy rate of about 43.13\%. The differences between contextualization and decontextualization were not significant for Gemini across different visualization types. In addition, we examined whether the accuracy rate for answering decontextualized questions without visualization would be closer to the rate when every choice is evenly and randomly selected. Overall, for all tasks in \gpt\ and the majority of tasks in \gemini, the accuracy rates were closer to the rate when choices were randomly picked. This suggests that for decontextualized questions, LLMs are more likely to guess rather than rely on their previous knowledge. We reported the detailed results in \autoref{app:follow_up_study_results}. 

In summary, in response to \textbf{RQ2.2}, it appears that contextualized visualizations tend to constrain LLMs to rely more on their pre-existing knowledge when answering questions. However, further research is needed to fully understand this relationship.

\vspace{-0.2em}
\section{Costs Differences between LLMs and Humans}
\edit{Though current LLMs cannot yet replace humans based on our results, we investigated} \redsout{Since the primary advantage of LLMs over human studies is the cost savings, we investigated}\textbf{RQ3}: \textit{What are the cost differences between LLMs and humans in interpreting visualizations and answering related questions?} \edit{to explore how the costs might differ if LLMs could accurately interpret visualizations in the similar level of humans in the future. It is not necessary for LLMs to perfectly interpret visualizations and answer questions, as they could still serve as an initial evaluation tool. Since cost efficiency is the primary advantage of LLMs over human-based studies, understanding those potential savings is crucial for future research planning.} We calculated the time and money spent on the experiments, comparing them with the expected costs of recruiting human participants to assess the extent of these differences. For \gpt, the API costs around \$0.01 for reading a visualization and answering the related question and \$0.001 for the same question without visualization. The \gemini\ API at the time of writing is free to use, so in total, excluding our pilot study and testing, we spent approximately \$160 for the formal study. In the test tryout of VLAT~\cite{Lee:2017:VDV}, each question was limited to 25 seconds for participants to answer. With 53 questions, it could take about 13 minutes to finish. Crowdsourcing platforms like Prolific have a minimum reward rate of \$8 per hour, resulting in about \$1.80 for each person finishing 53 questions. With \gpt, it took about \$0.53, and with \gemini, nothing. In general, using LLMs to analyze visualizations is much more economical.
In terms of time consumption, we found that \gpt's analysis time mainly depended on the API traffic. The most time-consuming trial took surprisingly 10,018 seconds for one question without visualization. When \gpt\ was asked about the estimated time needed for analyzing a visualization, it replied that it may take a few seconds. Based on this, we estimated that LLMs can typically analyze each of our visualizations within 100 seconds. Therefore, to offer a more concise overview, we filtered out trails that exceeded 100 seconds, considering them outliers. When running Experiment 1, the processing time of \gpt\ was faster for the initial requests, while the middle and final requests experienced slower processing time. We visualized detailed consumed time changes in Appendix~\ref{app:time_analysis}. On average, \gpt\ took 14.48 seconds to read and answer one visualization. \gemini\ remained relatively stable throughout the process. It took an average of 3.01 seconds to read a visualization and provide an answer. These findings suggest that, compared with the maximum answering time for each question set for humans (25 seconds), both LLMs are significantly faster, and \gemini\ is notably faster than \gpt. However, since we tested each question in a separate session, it is unclear whether testing LLMs in the manner of VLAT, which requires the model to simulate a person answering all 53 questions in one session, would alter the time taken.
As a whole, LLMs can be cost-effective and time-efficient in reading and understanding visualizations. However, their limited visualization literacy suggests a trade-off between using LLMs for evaluating visualizations and recruiting humans.

\vspace{-0.2em}
\section{Discussion}
\label{sec:discussion}
\vspace{-0.2em}
Visualization literacy is the foundation for evaluating the readability of visualizations~\cite{Cabouat:2024:PPC}. The aim of this paper is to propose a \edit{template} to evaluate whether current LLMs have visualization literacy and whether future LLMs can attain it to evaluate visualizations.
Our experiments revealed that the LLMs we used currently lack sufficient visualization literacy. We advocate for fine-tuning LLMs to enhance their performance in visualization-related tasks. The current GPT performed relatively well in tasks such as \textit{Find Correlations Trend} and \textit{Make Comparisons} in \textit{scatterplot} when compared to humans. In contrast, neither LLMs were reliable in reading \textit{pie chart} and \textit{histogram}. Given these performance differences, it is advisable to test both models to interpret the data values, and if a model can accurately read the data, we can adopt it for the initial evaluation of visualizations. \edit{The findings from another empirical evaluation study~\cite{Bendeck:2024:EEG}, which directly used VLAT for testing, suggest that \gpt\ performs well in tasks such as \textit{Finding Extremum}. However, our results show the opposite. This discrepancy may be due to LLMs leveraging prior knowledge from previous VLAT questions and answering based on other data resources. In the future, with advancements in LLM technology, we recommend that visualization experts adopt our methodology as an initial step to assess whether LLMs are capable of evaluating visualizations. This involves evaluating their visualization literacy by assessing new visualizations and comparing their performance to human results. With improved visualization literacy,} LLMs could become invaluable for the initial evaluation of new visualizations and systems. This approach could offer a cost-effective alternative to recruiting individuals on crowdsourcing platforms like Prolific. They could also potentially educate individuals about visualizations by providing design advice and explaining concepts using various types of charts. Also, we advocate for \textbf{a broader exploration of literacy beyond visualization}, as it remains unclear whether LLMs derive their answers solely from the intended questions or from their knowledge base. For instance, it is crucial to investigate LLMs' cross-modal understanding capability, \eg, integrating text and images. Also, exploring the textual or pictorial context understanding beyond the immediate modal, \eg, inferring hidden information or establishing connections between multiple pieces of information, can provide valuable insights for future usage.

An interesting insight we discovered is that \textbf{LLMs are influenced by factual accuracy}, or the ``real truth.'' Here, the ``real truth'' refers to questions that are specific and have corresponding datasets available online. For instance, information like unemployment rates for specific locations over the years or certain political facts are readily accessible and likely stored in LLMs' databases. This background information can cause LLMs to adhere to their learned knowledge, even if the visualizations suggest a different answer. This phenomenon is particularly pronounced in our experiments, where no choices were present for each question. This can also explain the good performance of LLMs in visualization reading in previous research to some extent. Although the visualizations were designed to reflect real scenarios and help people detect patterns efficiently, LLMs with biased pre-existing knowledge may fail to fully demonstrate the effectiveness of the visualizations themselves. Previous applications, \eg, \cite{Choe:2024:EDL}, have used real datasets to guide LLMs in helping people understand visualizations. However, our results suggest that LLMs may provide suggestions based on their pre-existing knowledge rather than accurately interpreting visualizations. Given that their knowledge is not updated in real-time, they are likely to fail to assist users in obtaining correct answers from visualizations encoded by new data. Therefore, current LLMs cannot replace humans in visualization tasks. Instead of merely trying to mitigate these limitations, we could potentially make use of this property when LLMs are evaluating visualizations. For example, LLMs can assess the data for potential fraud. With this capacity, LLMs can serve as a reminder to prompt visualization designers to reflect on the data source they used.

While current models are not yet prepared to replace humans in visualization evaluation, \textbf{they can be experimented with in various scenarios using diverse models}, especially considering their cost efficiency. For instance, designers could choose to employ \gpt\ to analyze their designed \textit{line charts}, \textit{histograms}, \textit{choropleth maps}, and \textit{treemaps}, while utilizing \gemini\ for testing their \textit{100\% stacked bar charts}, \textit{pie charts}, \textit{bubble charts}, and \textit{stacked area charts}. If the results are relatively poor, it probably indicates a design flaw. For an individual attempting a random visualization task, they could use \gpt\ to \textit{Find Correlations Trend} and \textit{Identify the Hierarchical Structure}, while employing \gemini\ for the other types of tasks. Moreover, removing the context of visualizations and adopting \gpt\ may be beneficial to enhance the interpretation performance. However, to provide more precise recommendations, further data collection involving a variety of visualizations and tasks is necessary to verify these approaches.






We also discovered that \textbf{LLMs are prone to overalignment}, a phenomenon where models exhibit an exaggerated tendency towards safety, significantly impacting their overall usefulness and reliability. For instance, Llama2-7b~\cite{Touvron:2023:Llama} demonstrated a 57\% rate of refusal to engage with innocuous prompts, indicating an overly cautious approach. In our study, this behavior was manifested primarily by the determined range without choices questions, where models provided overly broad or conservative responses to ensure technical correctness. For example, when answering the question about oil prices in 2015, which hovered around \$35 to \$65 a barrel, the model might give a range spanning the full extent of the y-axis or even beyond, sacrificing specificity for safety. While this cautiousness is aimed at avoiding incorrect or potentially harmful responses, it often results in outputs that are safe but lack utility or informative value. To address overalignment, a balance must be struck between factual accuracy and providing actionable information. This may require implementing more sophisticated training approaches that help LLMs grasp the subtleties of user queries and deliver responses that are both secure and useful.


\vspace{-0.5em}
\section{Limitations and Future Work}
\vspace{-0.2em}
Since LLMs are black boxes, there are various ways to test them, and different approaches may yield different performances. One limitation of our work is that we adopted a specific method to test their visualization literacy, and exploring other methods could be beneficial in the future. In our experiments, we tested each question in a separate session to minimize the influence of questions on each other, while VLAT asked participants to answer all 53 questions at once. Future work may involve having LLMs answer all 53 shuffled questions in one session. Also, we structured our prompts with questions preceding visualizations. Future studies can explore whether reversing this order has any effect.
Moreover, while we limited open-ended question responses to one word for ease of data analysis given the large volume of over 25,000 trials, this constraint may limit the performances of LLMs~\cite{Strobelt:2022:IVP}. In the future, we could focus on a relatively small number of trials and explore alternative prompting methods, such as Chain-of-Thought Prompting~\cite{Wei:2022:CTP}, to evaluate response accuracy and generation process. Another limitation of our work is that while we assessed the visualization literacy of LLMs, we did not assess how LLMs interact with data visualizations. Exploring interaction could expand the use of LLMs in reading visualizations. Another question that remains to be explored is whether LLMs can effectively analyze and provide insights when dealing with complex visualizations, such as dashboards.

Based on our experimental findings, there are several areas for future research. Our results indicate that, in the majority of cases, LLMs were affected by their pre-existing knowledge when answering questions. Therefore, we plan to investigate the impact of changing the prompt format. For instance, prompting LLMs with ``Please respond only using the information from the visualization, selecting the letter corresponding to the best option...'' could potentially lead to different outputs. In our follow-up study, we anonymized proper nouns, \eg, companies, countries, and political parties. However, we realized that there are different levels of decontextualization that could be influential in diverse ways. For instance, removing the specific category terms, such as ``Company,'' could prevent LLMs from searching for specific company names in the data. Additionally, using generic terms like `x' and `y' to represent data attributes could further anonymize the data. We are interested in exploring how these different levels of anonymization affect models' visualization literacy. We understand that while contexts are crucial for human understanding and engagement, they may not be as essential for LLMs in reading visualizations.
Furthermore, the scope of our research in this paper is to assess LLMs' visualization literacy using a heuristic approach following VLAT. In the future, we are interested in exploring their autonomous mode in reading and understanding visualizations. It would also be valuable to investigate why LLMs succeed or fail under specific conditions. Although it is challenging due to the opaque nature of LLMs, this direction can be further extended to investigate how LLMs evaluate visualizations in general and identify strategies to enhance their evaluative capabilities.


As LLMs continue to evolve, we understand that it is possible to have LLMs with advanced visualization literacy. Yet, we believe our results will remain informative and inspiring, and our methodologies could serve as a template for evaluating future models' visualization literacy. We urge practitioners to assess upgrading LLMs' visualization literacy regularly with our methodology so that we can further investigate the specific approaches to use them in visualization evaluation. 


\section*{Supplemental Material Pointers}

We provide the codes to generate visualizations, scripts to interact with LLMs, and scripts to analyse experimental results at \href{https://github.com/VADERASU/llm4viz-experiments}{\texttt{github.com/VADERASU/llm4viz-experiments}}. We also released our visualization used in the studies, experiential results, and figures in the manuscript at \href{https://osf.io/wcb5g/}{\texttt{osf.io/wcb5g}}.

\bibliographystyle{abbrv-doi-hyperref}
\bibliography{template}

\clearpage
\appendix 

\section{Results of LLMs' Visualization Literacy}
\subsection{Variables and Interactions}\label{app:varInterCalc}
For the logistic regression model, recall that we denote $\textbf{x}_k$ where $k \in \{1,2,3,4\}$ as the variables corresponding to the $k$-way interaction variables (see \autoref{eq:model}). More explicitly, let $V$ be the set of visualization types, $T$ be the set of task types, $L$ be the set of LLMs, and $P$ be the set of visualization presence.  Then, let $D$ be the family of these dimension sets such that $\{V,T,L,P\} \equiv D$. Let 
$D_k=\begin{pmatrix}
    D\\
    k\\
\end{pmatrix}$ denote the $k$-subset of $D$ where $|D_k|=m$, and $D_{k,j}$ denote the $j$th element in $D_k$ where $j \in \{1,\dots,m\}$ for some ordering. For example, $D_{2,1}$ could be $\{V,T\}$.  Then, let $C(D_{k,j})=\prod\limits_{S \in D_{k,j}}S$ be the Cartesian product of $D_{k,j}$, and $C(D_{k,j})_l$ denote the $l$th tuple in $C(D_{k,j})$ where $l \in \{1,\dots,n_j\}$ and $n_j=|C(D_{k,j})|$.  For instance, $C(D_{2,1})=\prod\limits_{S \in \{V,T\}}S=V \times T$, which describes all two-way interactions between visualization and task types. This includes visualization/task interactions that were not tested, \eg, Line Chart and Find Anomalies, though in reality, we did not include them in our model\footnote{The same thing can be accomplished by including these variables but setting their coefficients to zero}. Then $\textbf{x}_k$ can be expressed as the following:

\begin{equation}
    \textbf{x}_k=
    \begin{bmatrix}
        \prod\limits_{i \in C(D_{k,1})_1}x_i\\
        \vdots\\
        \prod\limits_{i \in C(D_{k,1})_{n_1}}x_i\\
        \vdots\\
        \prod\limits_{i \in C(D_{k,m})_1}x_i\\
        \vdots\\
        \prod\limits_{i \in C(D_{k,m})_{n_m}}x_i\\
    \end{bmatrix}
\end{equation}

\noindent where $\textbf{x}_k$ is a $(n_1+\dots+n_m) \times 1$ vector, and $x_i$ is a binary variable that determines whether variable $i$ is included where $i$ is an element of the $C(D_{k,j})_l$ tuple. With 12 visualization types, 8 task types, 2 LLMs, a binary variable for visualization presence, and 49 unique visualization and task type interactions, the number of variables with their interactions is calculated as follows:\\

\noindent One-way: $12+8+2+2=24$\\
Two-way: $49+(12 \times 2)+(12 \times 2)+(8 \times 2)+(8 \times 2)+(2 \times 2)=133$\\
Three-way: $(49 \times 2)+(49 \times 2)+(12 \times 2 \times 2)+(8 \times 2 \times 2)=276$\\
Four-way: $49 \times 2 \times 2=196$\\

\noindent Thus, a total of $24+133+276+196=629$ variables and interactions were used for our logistic regression model.

\subsection{Hyperparameter Tuning}\label{sec:paramSelect}
Using scikit-learn's \texttt{LogisticRegression} function,  we measured five metrics to evaluate the best model parameters: \texttt{Average Precision Score (APS)}, \texttt{Area under the Precision-Recall Curve (AUPRC)}, \texttt{Area under the Receiver Operator Curve (AUCROC)}, \texttt{F1 score}, and \texttt{accuracy}.  Almost all of the metrics were calculated using scikit-learn's metrics API.  For the area under the curve calculations, the $x$ and $y$ values were found first using built-in curve estimators and then calculated using scikit-learn's \texttt{AUC} function.

Based on the \texttt{LogisticRegression} documentation, we performed parameter selection on various penalties, solvers, and regularization values.  \autoref{tab:logRegParam} shows which penalties and solvers were tested.  Besides the no penalty case, $cVal \in \{0.001,0.01,0.1,1,10,100,1000\}$ where $cVal$ is the regularization value.  This corresponds to the \texttt{C} variable in the sci-kit learn documentation.  For elastic net, $\ell_1 \text{ratio} \in \left\{0,\frac{1}{9},\frac{2}{9},\frac{1}{3},\frac{4}{9},\frac{5}{9},\frac{2}{3},\frac{7}{9},\frac{8}{9},1\right\}$ where $\ell_1 \text{ratio}$ refers to the weight for the LASSO (and subsequently the Ridge) penalty.  This corresponds to the \texttt{l1\_ratio} in the sci-kit learn documentation.  In total, there are $(1 \times 7)+(5 \times 7)+(5)+(1 \times 7 \times 10)=117$ unique parameter combinations.  For each of these parameter combinations, we performed ten ten-fold cross-validations and recorded their accuracy scores on the test sets, totaling 11,700 data points.

\begin{table}[h]
    \centering
    \caption{The table of penalties and solvers used for parameter selection for logistic regression}
    \begin{tabu}{l|l}
    \hline
        \textbf{Penalty}            &   \textbf{Solver}\\
        \hline
        $l_1$ norm                  &   liblinear\\
        \hline
        \multirow{5}*{$l_2$ norm}   &   lbfgs\\
                                    &   liblinear\\
                                    &   newton-cg\\
                                    &   newton-cholesky\\
                                    &   sag\\
        \hline
        \multirow{5}*{No Penalty}   &   lbfgs\\
                                    &   liblinear\\
                                    &   newton-cg\\
                                    &   newton-cholesky\\
                                    &   sag\\
        \hline
        elastic net                 &   saga\\
        \hline
    \end{tabu}
    \label{tab:logRegParam}
\end{table}

A total of 9,185 questions were answered correctly out of the 25,440 samples ($\approx 36.1\%$ correct), meaning our data was imbalanced.  Since AUPRC is robust to data imbalance, this was the primary metric used to determine the best logistic regression model.  Based on boxplots of AUPRC for the various penalties (see supplementary materials), the no penalty model performed the best with the $\ell_1$, $\ell_2$, and elastic net penalties all performing on par with $C \geq 10$ (i.e., relatively small regularization penalty coefficients
).  In fact, all of the metrics seemed to follow the same convergence structure for all of the penalties.  Further, all of the solvers for the no-penalty model appeared to converge to the same results, so the default lbfgs solver was used.

\subsection{Bootstrapping}\label{app:bootstrapping}
We sampled 25,440 data points with replacement from experiments 1 and 2 and fitted a logistic regression model using the best hyperparameters 1,000 times. Each time we fit a bootstrapped model, we stored the 629 coefficients that resulted from that model.  We then analyzed the coefficients from a distribution perspective to see if they were different from zero, which allowed us to answer \textbf{RQ1.1} and test hypotheses \textbf{H1}, \textbf{H2}, and \textbf{H3}.  In addition, each bootstrapped model allowed us to calculate the probability of specific visualization/task/LLM/visualization presence combinations and answer \textbf{RQ1.1} and \textbf{RQ1.2}, hypotheses \textbf{H1} and \textbf{H4} especially. \autoref{tab:bootstrap} shows the structure of how we stored the coefficient data and how the probabilities were associated with each bootstrap model.

\begin{table}[h]
    \centering
    \caption{The table demonstrating the structure of our bootstrap methodology. The main model is based on all of the data from experiments 1 and 2, while the bootstrap models are based on sampling the same data with replacement.}
    \begin{tabu}{l|llll|l}
        \multirow{2}*{\textbf{Models}}  &   \multicolumn{4}{c|}{\textbf{Coefficients}}                              &   \multirow{2}*{\textbf{Probabilities}}\\
        \cline{2-5}
                                        &   \textbf{Intercept}      &   \textbf{Coef. 1}        &  \textbf{\dots}   &  \textbf{Coef. 629}       &\\
        \hline
        \textbf{Main}                   &   $\hat{\beta}_0$         &   $\hat{\beta}_1$         &   $\dots$         &   $\hat{\beta}_{629}$     &   $\hat{P}(y=1|\textbf{x}),\dots$\\
        \hline
        \textbf{Bootstrap 1}            &   $\hat{\beta}_{0,1}$     &   $\hat{\beta}_{1,1}$     &   $\dots$         &   $\hat{\beta}_{629,1}$   &   $\hat{P}_{1}(y=1|\textbf{x}),\dots$\\
        $\vdots$                        &   $\vdots$                &   $\vdots$                &   $\ddots$        &   $\vdots$                &   $\vdots$\\
        \textbf{Bootstrap 1000}         &   $\hat{\beta}_{0,1000}$  &   $\hat{\beta}_{1,1000}$  &   $\dots$         &   $\hat{\beta}_{629,1000}$&   $\hat{P}_{1000}(y=1|\textbf{x}),\dots$\\
        \hline
    \end{tabu}
    \label{tab:bootstrap}
\end{table}

\subsection{Bootstrapped Results}\label{app:bootstrapped}
\subsubsection{Coefficients}\label{app:coefficients}
\begin{table}[!h]
    \centering
    \caption{Statistically insignificant coefficients for logistic regression model.}
    \begin{tabu}{l|l|l|l}
        \hline
        \textbf{Visualization}      &\textbf{Task}                  &\textbf{Vis. Presence} &\textbf{LLM}\\
        \hline
        \multirow{4}*{Line chart}   &\multirow{3}*{Find Extremum}   &Vis.                   &Gemini\\
        \cline{3-4}
                                    &                               &\multirow{2}*{No Vis.} &Gemini\\
        \cline{4-4}
                                    &                               &                       &Both\\
        \cline{2-4}
                                    &Make Comperison                &No. Vis                &Both\\
        \hline
        100\% Stacked Bar Chart     &Find Extremum                  &No Vis.                &Gemini\\
        \hline
        Scatterplot                 &Retrieve Value                 &No Vis.                &GPT-4\\
        \hline
    \end{tabu}
    \label{tab:insigCoefs}
\end{table}

\begin{figure}[h]
    \centering
    \includegraphics[width=0.49\textwidth]{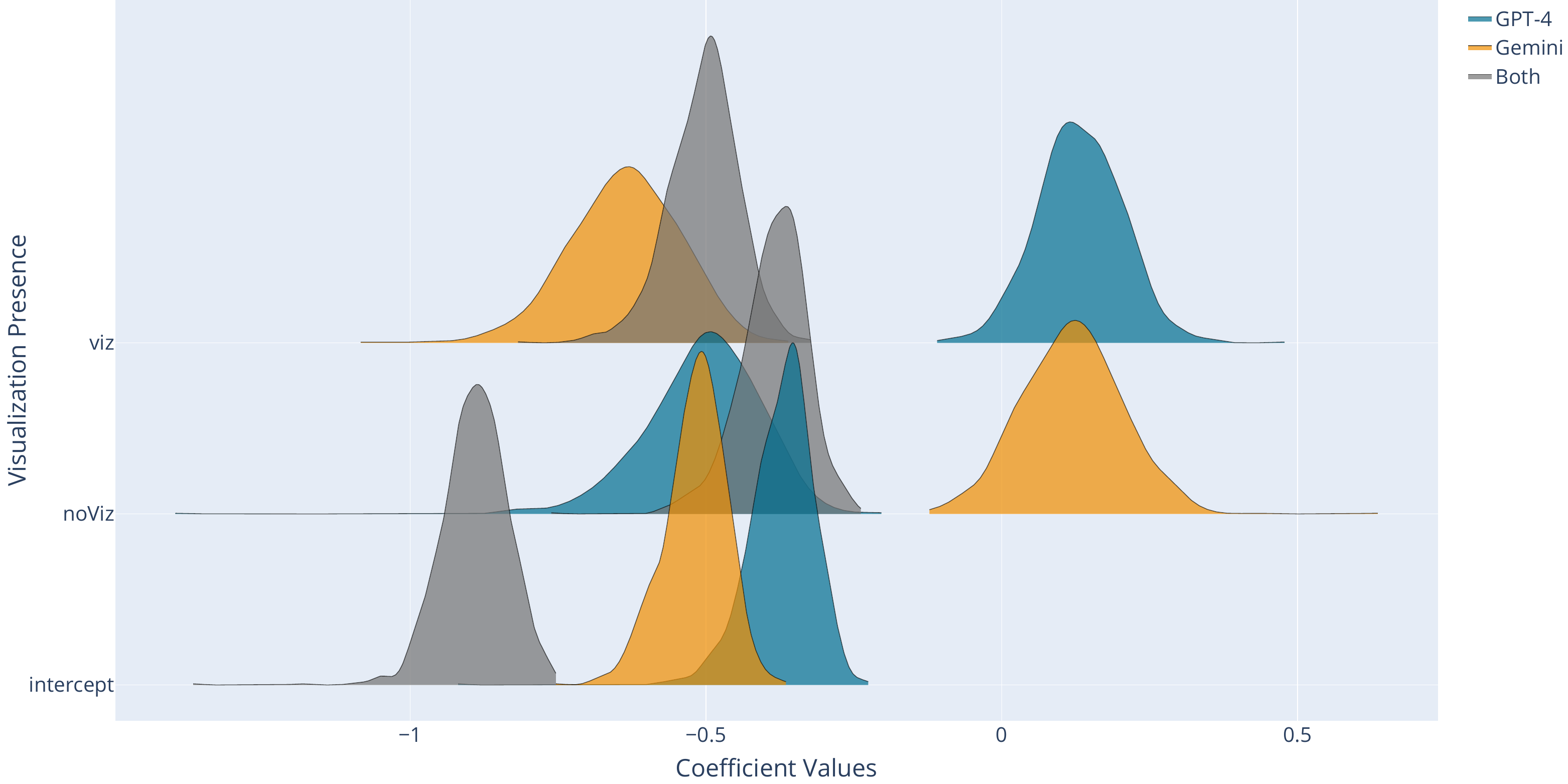}
    \caption{Ridge plot of the bootstrapped general coefficients from the logistic regression.}
    \label{fig:genCoef}
\end{figure}

As previously mentioned in Section~\ref{sec:coefResults}, all the coefficients were found to be statistically significant except for 6, which we list in \autoref{tab:insigCoefs}. The general coefficients (\ie, LLM and visualization presence variables and interactions) are visualized in \autoref{fig:genCoef} as a ridge plot and were found to be statistically significant (see \autoref{tab:genCoefResults}). A table of the mean visualization coefficients can be found in \autoref{tab:vizCoefResults}, and a similar table for the mean task coefficients can be found in \autoref{tab:taskCoefResults}.  The visualization/task interaction coefficients when the visualization is present are visualized in \autoref{fig:chartTaskInterCoef} as a ridge plot with all of the mean values found in \autoref{tab:vizTaskCoefResults}.

\subsubsection{\gpt\ vs. \gemini\ Probability Differences}\label{app:bootstrapLLM}
\autoref{tab:llmDiffResults} includes the two-sided test results for significant differences between GPT-4 and Gemini when the visualization is present. This table includes the visualization type, task type, whether the beta difference distribution or the Wilcoxon signed-rank test was used, the corresponding p-value, and the 95\% confidence interval.  It shows that of the 49 visualization/task interactions, 17 showed GPT-4 outperforming Gemini, 25 showed Gemini performing better, and 7 showed no difference.  The largest difference in GPT-4's favor was \textit{Identify Hierarchical Structure} for \textit{treemap} ($95\% \text{ CI} \approx (1.000,1.000)$) while the largest difference for Gemini was \textit{Make Comparisons} for \textit{100\% stacked bar chart} ($95\% \text{ CI} \approx (-0.8895,-0.7170)$).

\subsubsection{Visualization Presence Probability Differences}\label{app:visdiffResults}
\autoref{tab:vizDiffResults} includes the one-sided test results for significant differences between visualization presence and absence. This table includes the LLM, visualization type, task type, whether the beta difference distribution or the Wilcoxon signed-rank test was used, the corresponding p-value, and the 95\% confidence interval.  18 out of the 49 visualization/task interactions were statistically significant for GPT-4, and 20 were significant for Gemini.  The largest difference for GPT-4 was \textit{Find Correlation/Trend} for \textit{line chart} ($95\% \text{ LB} \approx 0.8161$), while the largest difference for Gemini was \textit{Retrieve Value} for \textit{100\% stacked bar chart} ($95\% \text{ LB} \approx 0.5573$).

\subsection{Time Analysis}\label{app:time_analysis}
\begin{figure}[h]
    \centering
    \begin{subfigure}{.24\textwidth}
        \includegraphics[width=\textwidth]{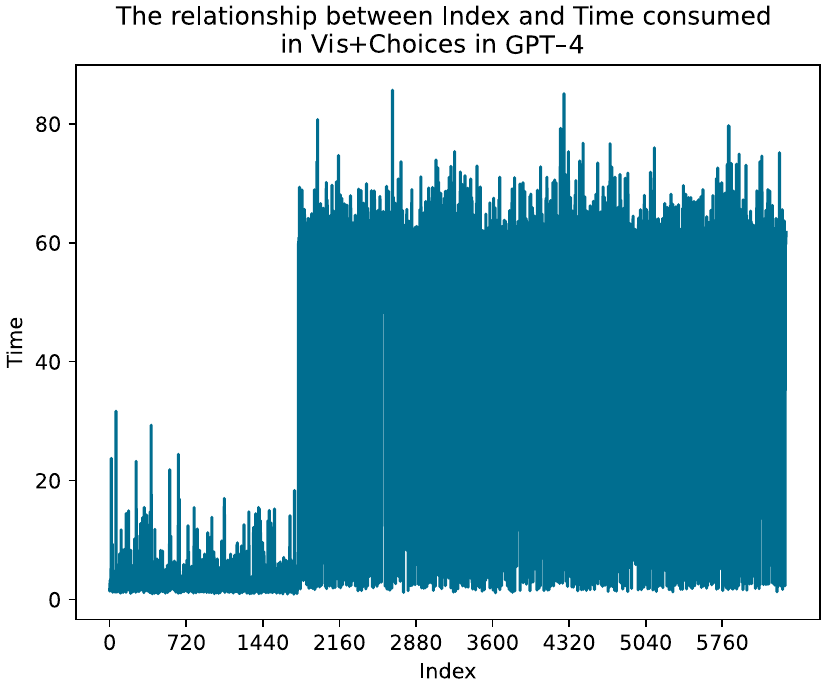}
        \caption{GPT-4 for visualization with choices}
        \label{fig:time_GPT-4_choices}
    \end{subfigure}
    \begin{subfigure}{.24\textwidth}
        \includegraphics[width=\textwidth]{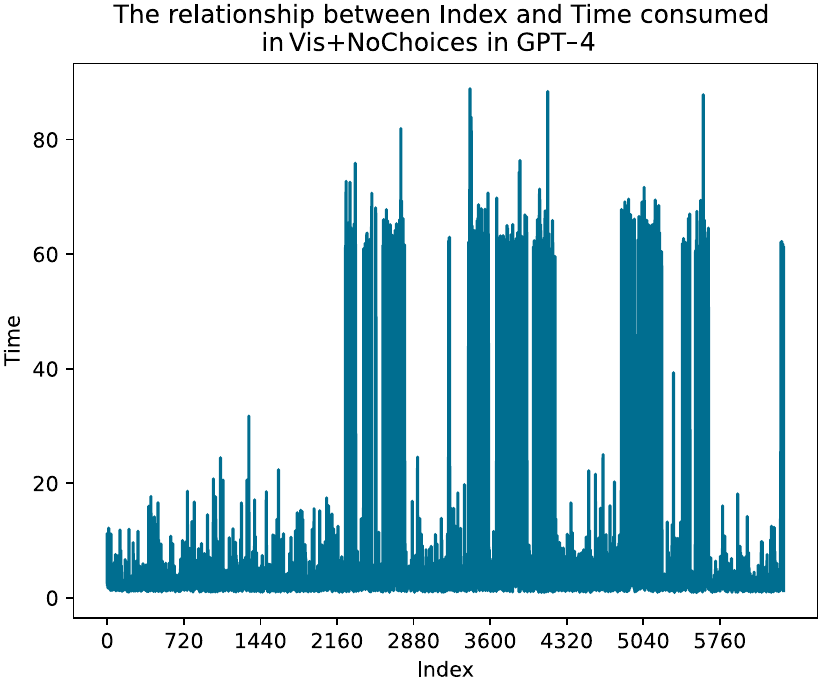}
        \caption{GPT-4 for visualization without choices}
        \label{fig:time_GPT-4_nochoices}
    \end{subfigure}
    \begin{subfigure}{.24\textwidth}
        \includegraphics[width=\textwidth]{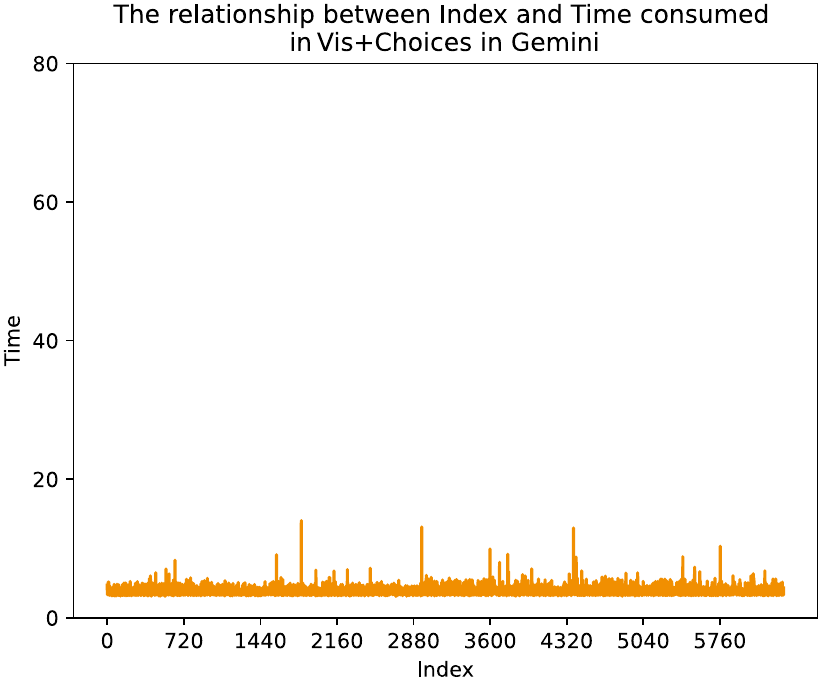}
        \caption{Gemini for visualization with choices}
        \label{fig:time_gemini_choices}
    \end{subfigure}
    \begin{subfigure}{.24\textwidth}
        \includegraphics[width=\textwidth]{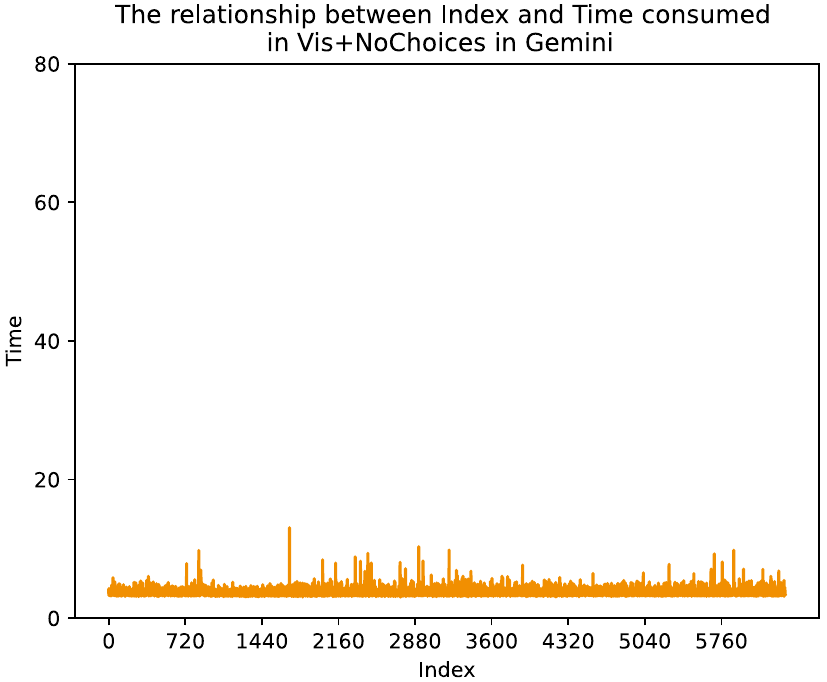}
        \caption{Gemini for visualization without choices}
        \label{fig:time_gemini_nochoices}
    \end{subfigure}
    \caption{The time consumption of two LLMs for each visualization in different conditions over time.}
    \label{fig:time_models_vis}
\end{figure}

We visualized the detailed time consumption for two LLMs, as shown in \autoref{fig:time_models_vis}. GPT-4 generally took longer than Gemini in both cases (with or without choices). Additionally, GPT-4 exhibited relatively unstable response times compared to Gemini when subjected to a continuous series of requests. 

\section{Experiments without Choices}\label{app:analysis_nochoices}
In this section, we report the detailed analysis methods and the results of Experiments 3 and 4.


\subsection{Numerical Range Analysis and Results}
\label{app:numerical_range_ana}

\begin{figure*}[h]
    \centering
    \includegraphics[width=\linewidth]{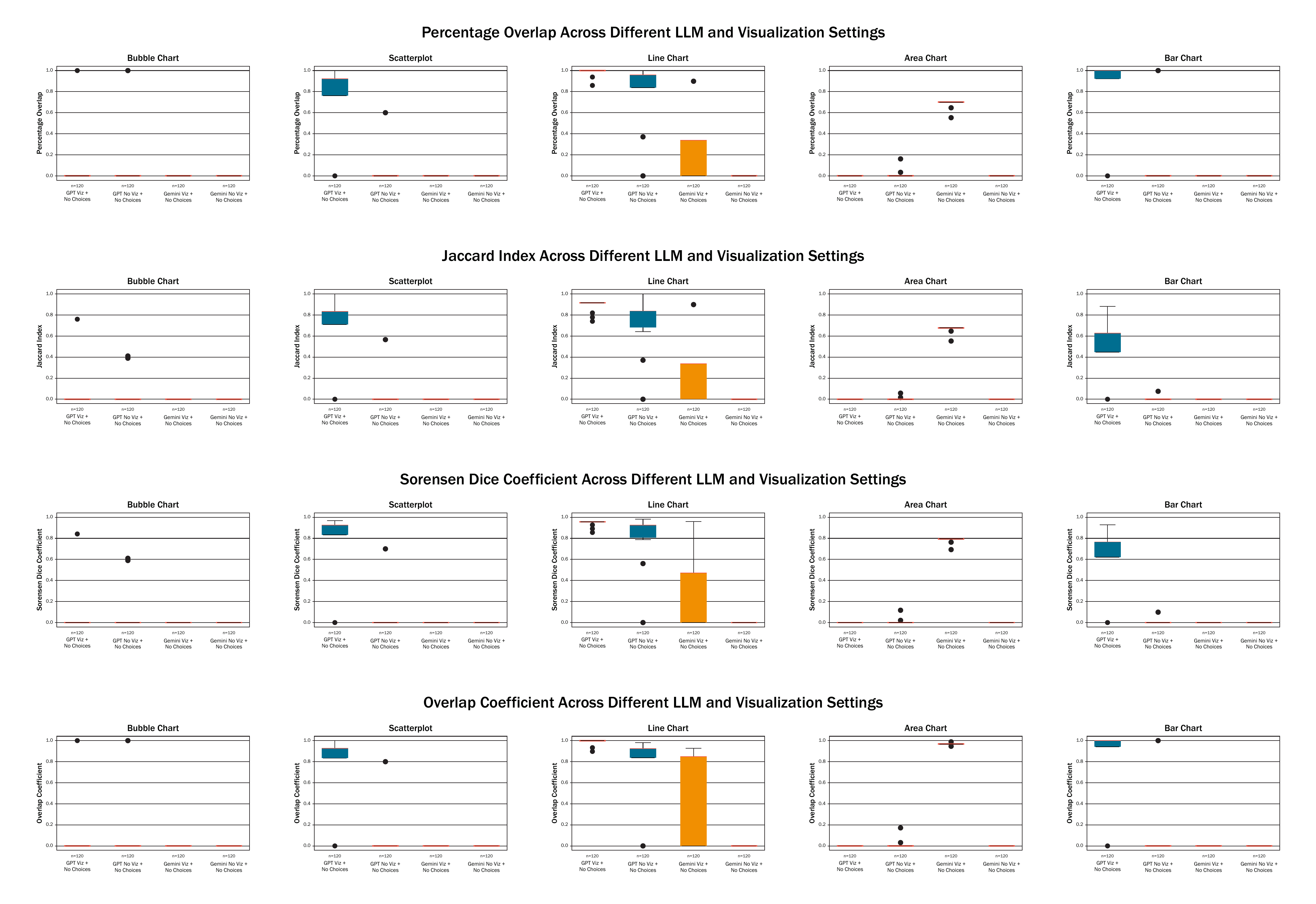}\vspace{-1ex}
    \caption{Box plots showing the overlap between LLMs' responses and the correct answers for questions determining the range. The plots depict the distribution of four metrics we used to measure this overlap.}
    \label{fig:determine_range_no_choices}
\end{figure*}

In this section, we describe the detailed metrics we used to gain a more comprehensive picture of the overlap. We visualized the results of these four metrics in \autoref{fig:determine_range_no_choices}. 

\textit{Percentage Overlap} is defined as
\begin{equation}
\label{eq:percentage_overlap}
    \text{Overlap} = 
\frac{\text{Size of Overlap between A and B}}{\text{Total Size}}
\end{equation}

Although calculating percentage overlap is straightforward, it fails to adequately reflect the usefulness of the response. When $A$ is a subset of $B$ or vice versa, the percentage overlap is, by definition, 100\%. However, this metric may become impractical if the sizes of the ranges are significantly different. For example, if the correct range is 0-10 and the model's response is 0-1000, this result is virtually useless. Similarly, an accurate range of 0-1000 with a model response of 0-10 is also problematic. Despite the model capturing only a minuscule portion of the full range, the percentage overlap misleadingly remains at 100\%.

\textit{Jaccard Index} is defined as~\cite{Costa:2021:Jaccard}.
\begin{equation}
\label{eq:Jaccard}
J(A, B) = \frac{|A \cap B|}{|A \cup B|}
\end{equation}

When the Jaccard Index is applied to intervals, the values can range from 0 (no overlap) to 1 (identical intervals). For example, if interval A spans from 1 to 4 and interval B from 3 to 6, the intersection is from 3 to 4, and their union is from 1 to 6, resulting in a Jaccard Index of $\frac{1}{5}$. This index does compare the overall similarity between intervals but might not fully represent the proportionality of overlap when intervals vary widely in length.

\textit{Sørensen-Dice Coefficient} is defined as~\cite{Jackson:1989:Similarity}.
\begin{equation}
\label{eq:Sorensen}
D(A, B) = \frac{2|A \cap B|}{|A| + |B|}
\end{equation}

This metric results in a measure that expresses the overlap in relation to the total size of both intervals, with a value ranging from 0 (indicating no overlap) to 1 (indicating identical intervals). An in-depth example might involve Interval A spanning from 0 to 4 and Interval B from 2 to 6. The intersection, or shared portion, spans from 2 to 4, giving a length of 2. The total combined length of both intervals (from the start of A to the end of B) is 6. Plugging these into the formula gives a Sørensen-Dice Coefficient of $\frac{4}{10} = 0.4$, indicating a moderate level of similarity between the two intervals. Like the Jaccard Index, The Sørensen-Dice Coefficient is particularly useful in contexts where the balance between the shared portion and the total extent of intervals matters.

\textit{Overlap Coefficient} is defined as~\cite{Vijaymeena:2016:Survey}.
\begin{equation}
\label{eq:overlap_coefficient}
O(A, B) = \frac{|A \cap B|}{\min(|A|, |B|)}
\end{equation}

This metric provides a measure of how much one interval is contained within another by comparing the size of their intersection to the size of the smaller interval. The result is a value between 0 and 1, where 0 means there is no overlap and 1 indicates that the smaller interval is completely contained within the larger one. For a more detailed illustration, assume Interval A spans 1 to 3 (length 2) and Interval B spans 2 to 8 (length 6). Their intersection is from 2 to 3, with a length of 1. Since Interval A is smaller, the Overlap Coefficient is calculated as $\frac{1}{2} = 0.5$, indicating that half of Interval A overlaps with Interval B. This metric is most important when assessing the extent to which one dataset (represented by an interval) covers or is representative of another. It is highly relevant in scenarios where the complete capture or representation of the smaller interval by the larger one is critical, effectively measuring containment rather than just similarity.

\section{Details about Experiment 5}\label{app:follow_up_study_results}
\begin{figure*}[ht]
    \centering
    \includegraphics[width=\linewidth]{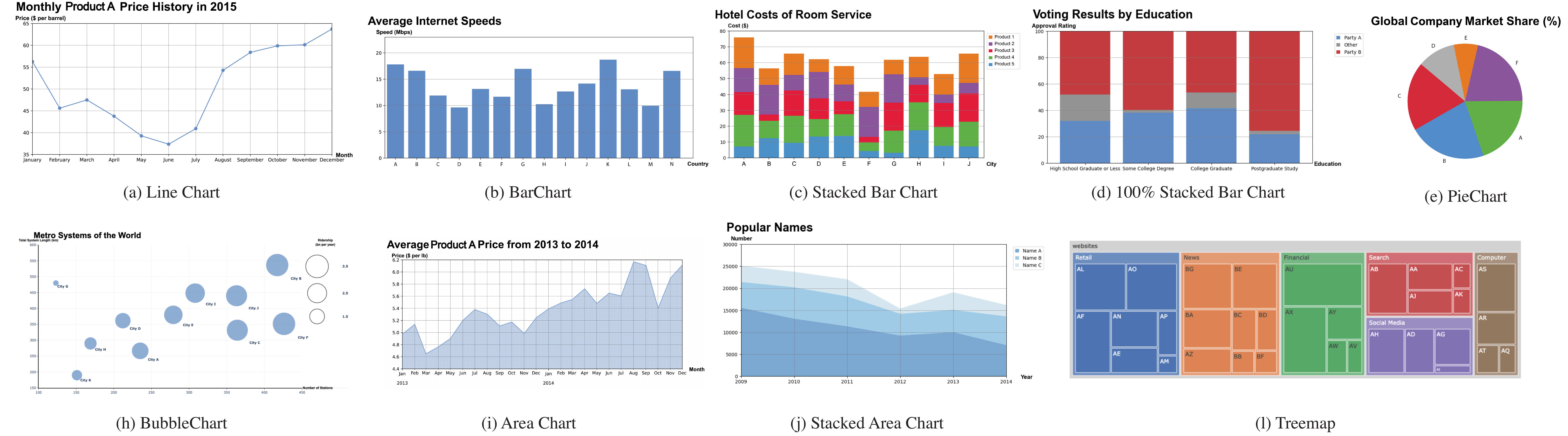}\vspace{-1ex}
    \caption{Eight decontextualized visualizations tested in our follow-up experiment.}
    \label{fig:visualizations_anon}
\end{figure*}

We reported the values of the follow-up study, where anonymized visualizations were investigated (see \autoref{fig:visualizations_anon}), compared with the results of Experiment 1 in \autoref{tab:follow_up_experiment_result} qualitatively. With decontextualized visualizations, GPT-4 tended to improve the performance, while the differences were not significant for Gemini.


\begin{table*}[h]
\centering
\caption{The comparisons between the accuracy rates for the results of GPT-4 and Gemini in Experiment 1 and the follow-up experiment. We color-encoded LLMs' results of reading visualizations in the follow-up experiment: green for much better than the performance of Experiment 1 (more than 0.05 higher), yellow for values that were close to those of Experiment 1 (higher or lower in 0.05), and red for values much worse than those of Experiment 1. We also color-encoded the results of answering questions without visualizations. If the results in the follow-up study are close to the random-choice possibility, we mark it in green. Otherwise, we mark it in red.}
\begin{tabu} {lllcccccccc}
\hline
\multirow{3}{*}{Visualization} & \multirow{3}{*}{Task} & \multicolumn{4}{c}{With Visualization} & \multicolumn{4}{c}{Without Visualization} & \multirow{3}{*}{Random} \\
        &  & \multicolumn{2}{c}{Experiment 1} & \multicolumn{2}{c}{Follow-up} & \multicolumn{2}{c}{Experiment 1} & \multicolumn{2}{c}{Follow-up} \\
        &  & GPT-4 & Gemini & GPT-4 & Gemini & GPT-4 & Gemini & GPT-4 & Gemini & \\
    \hline
    \multirow{5}{*}{Line Chart} & Retrieve Value & 0.56 & 0.25 & \worse{0.31} & \close{0.25} & 0.43 & 0.28 & \better{0.32} & \close{0.22} & 0.25 \\
                                & Find Extremum & 0.02 & 0.22 & \better{0.36} & \close{0.25} & 0.03 & 0.22 & \close{0.03} & \better{0.24} & 0.25 \\
                                & Determine Range & 0.23 & 0.23 & \worse{0.17} &\worse{0.16} & 0.52 & 0.25 & \better{0.42} & \worse{0.22} & 0.25 \\
                                & Find Correlation/Trend & 0.87 & 0.42 & \better{1.00} & \better{0.83} & 0.00 & 0.05 & \better{0.01} & \better{0.34} & 0.33 \\
                                & Make Comparisons & 0.87 & 0.28 & \worse{0.41} & \close{0.26} & 0.23 & 0.20 & \better{0.26} & \better{0.24} & 0.25 \\
    \hline
    \multirow{4}{*}{Bar Chart}  & Retrieve Value & 0.40 & 0.56 & \close{0.35} & \worse{0.48} & 0.26 & 0.58 & \worse{0.42} & \better{0.47} & 0.25 \\
                                & Find Extremum & 0.00 & 0.01 & \better{0.28} & \close{0.00} & 0.00 & 0.03 & \better{0.33} & \better{0.31} & 0.25 \\
                                & Determine Range & 0.29 & 0.68 & \better{0.48} & \worse{0.57} & 0.38 & 0.76 & \better{0.18} & \better{0.56} & 0.25 \\
                                & Make Comparisons & 0.20 & 0.13 & \better{0.26} & \better{0.18} & 0.33 & 0.25 & \better{0.28} & \worse{0.28} & 0.25 \\
    \hline
    \multirow{5}{*}{\makecell[l]{Stacked \\ Bar Chart}} 
                            & Retrieve Value (Absolute Value) & 0.23 & 0.25 & \worse{0.09} & \worse{0.13} & 0.36 & 0.24 & \worse{0.11} & \worse{0.12} & 0.25 \\
                            & Retrieve Value (Relative Value) & 0.04 & 0.23 & \better{0.14} & \close{0.21} & 0.23 & 0.35 & \worse{0.49} & \better{0.28} & 0.25 \\   
                            & Find Extremum & 0.17 & 0.05 & \better{0.31} & \better{0.21} & 0.46 & 0.02 & \better{0.38} & \better{0.43} & 0.25 \\
                            & Make Comparisons (Absolute Value) & 0.70 & 0.78 & \better{0.82} & \worse{0.02} & 0.88 & 0.24 & \better{0.53} & \worse{0.00} & 0.50 \\
                            & Make Comparisons (Relative Value) & 0.94 & 0.43 & \close{0.96} & \better{0.48} & 0.51 & 0.12 & \better{0.50} & \worse{0.01} & 0.50 \\
    \hline
    \multirow{3}{*}{\makecell[l]{100\% Stacked \\ Bar Chart}} 
                            & Retrieve Value (Absolute Value) & 0.00 & 0.79 & \better{0.19} & \close{0.83} & 0.00 & 0.15 & \better{0.19} & \better{0.28} & 0.25 \\
                            & Find Extremum (Relative Value) & 0.00 & 0.21 & \better{0.76} & \worse{0.18} & 0.00 & 0.03 & \close{0.53} & \close{0.04} & 0.25 \\
                            & Make Comparisons (Relative Value) & 0.15 & 0.98 & \worse{0.07} & \worse{0.71} & 0.10 & 0.87 & \better{0.48} & \worse{0.98} & 0.50 \\
    \hline
    \multirow{3}{*}{Pie Chart}
                            & Retrieve Value (Relative Value) & 0.00 & 0.36 & \better{0.28} & \better{0.69} & 0.00 & 0.06 & \better{0.29} & \better{0.20} & 0.25 \\
                            & Find Extremum (Relative Value) & 0.00 & 0.30 & \better{0.28} & \worse{0.05} & 0.97 & 0.90 & \close{0.03} & \better{0.23} & 0.25 \\
                            & Make Comparisons (Relative Value) & 0.00 & 0.43 & \better{0.43} & \better{0.99} & 0.00 & 0.02 & \better{0.50} & \close{0.98} & 0.50 \\
    \hline
    \multirow{4}{*}{Area Chart}
                            & Retrieve Value & 0.28 & 0.13 & \worse{0.21} & \close{0.09} & 0.26 & 0.16 & \close{0.27} & \close{0.17} & 0.25 \\
                            & Find Extremum & 0.57 & 0.73 & \better{0.78} & \close{0.73} & 0.54 & 0.37 & \close{0.55} & \better{0.25} & 0.25 \\
                            & Determine Range & 0.23 & 0.13 & \close{0.21} & \better{0.23} & 0.28 & 0.19 & \worse{0.13} & \better{0.28} & 0.25 \\
                            & Find Correlation/Trend & 1.00 & 1.00 & \worse{0.89} & \worse{0.94} & 1.00 & 0.68 & \better{0.58} & \better{0.33} & 0.33 \\
    \hline
    \multirow{6}{*}{\makecell[l]{Stacked \\ Area Chart}}
                            & Retrieve Value (Absolute Value) & 0.36 & 0.35 & \close{0.32} & \worse{0.30} & 0.42 & 0.28 & \better{0.27} & \close{0.23} & 0.25 \\
                            & Retrieve Value (Relative Value) & 0.17 & 0.30 & \better{0.54} & \close{0.33} & 0.01 & 0.33 & \better{0.27} & \better{0.22} & 0.25 \\
                            & Find Extremum & 0.00 & 0.01 & \better{0.14} & \close{0.00} & 0.00 & 0.03 & \better{0.21} & \close{0.03} & 0.25 \\
                            & Find Correlation/Trend & 0.00 & 0.72 & \better{0.42} & \better{1.00} & 0.00 & 0.17 & \better{0.14} & \better{0.33} & 0.33 \\
                            & Make Comparisons (Absolute Value) & 0.51 & 0.99 & \worse{0.08} & \worse{0.94} & 0.94 & 0.93 & \better{0.69} & \close{0.89} & 0.50 \\
                            & Make Comparisons (Relative Value) & 1.00 & 1.00 & \close{0.97} & \close{0.98} & 0.99 & 0.49 & \better{0.54} & \worse{0.00} & 0.50 \\
    \hline                            
    \multirow{7}{*}{Bubble Chart}
                            & Retrieve Value & 0.02 & 0.15 & \better{0.27} & \better{0.25} & 0.00 & 0.06 & \better{0.31} & \better{0.18} & 0.25 \\
                            & Find Extremum & 0.00 & 0.11 & \better{0.28} & \better{0.30} & 0.03 & 0.06 & \better{0.23} & \better{0.34} & 0.25 \\
                            & Determine Range & 0.25 & 0.26 & \better{0.32} & \close{0.23} & 0.02 & 0.26 & \close{0.03} & \worse{0.18} & 0.25 \\
                            & Find Anomalies & 0.53 & 0.06 & \better{0.61} & \better{0.36} & 0.20 & 0.26 & \worse{0.49} & \worse{0.38} & 0.25 \\
                            & Find Clusters & 0.38 & 0.48 & \worse{0.23} & \better{0.59} & 0.68 & 0.38 & \better{0.50} & \better{0.53} & 0.50 \\
                            & Find Correlation/Trend & 0.93 & 1.00 & \close{0.97} & \worse{0.88} & 1.00 & 0.97 & \close{1.00} & \close{1.00} & 0.50 \\
                            & Make Comparisons & 0.00 & 0.09 & \better{0.19} & \better{0.60} & 0.03 & 0.72 & \better{0.50} & \worse{0.84} & 0.50 \\
    \hline
    \multirow{3}{*}{Treemap}
                            & Find Extremum (Relative Value) & 0.01 & 0.00 & \better{0.58} & \better{0.17} & 0.00 & 0.01 & \better{0.29} & \better{0.14} & 0.25 \\  
                            & Make Comparison (Relative Value) & 0.03 & 0.53 & \close{0.03} & \better{0.88} & 0.46 & 0.90 & \close{0.43} & \better{0.82} & 0.50 \\
                            & Identify the Hierarchical Structure & 1.00 & 0.00 & \worse{0.85} & \close{0.00} & 0.75 & 0.19 & \better{0.53} & \worse{0.08} & 0.50 \\
    \hline
\end{tabu}
\label{tab:follow_up_experiment_result}
\end{table*}

\begin{table*}[h]
    \centering    
    \caption{Table of coefficients for LLMs and visualization presence interactions.  \textbf{Coef.} refers to the coefficient in the logistic model using results in experiments 1 and 2, while the \textbf{Mean Coef.} refers to the mean of the bootstrapped coefficients.  When \textbf{Normal?} is true, a one-sample t-test is used, while if it is false, the Wilcoxon signed-rank test is used.  Note that when the Wilcoxon signed-rank test is used, the bounds are estimated from the ECDF.  Thus, bounds calculated from the ECDF that contain zero indicate poor estimations.}
    \begin{tabu}{l|l|S[table-format=1.4e1]|S[table-format=1.4e1]|S[table-format=1.4e3]|l|S[table-format=1.4e1]|S[table-format=1.4e1]}
        \hline
        \textbf{LLM}            &\textbf{Vis. Presence} &\textbf{Coef.} &\textbf{Mean Coef} &\textbf{p-value}   &\textbf{Normal?}   &\textbf{95\% LB}   &\textbf{95\% UB}\\
        \hline
        \multirow{2}*{Both}     &Vis. Present           &-0.4737        &-0.5051            &3.326E-165         &False              &-0.6379            &-0.3967\\
                                &No Vis. Present        &-0.3860        &-0.3855            &3.326E-165         &False              &-0.5141            &-0.2833\\
        \hline
        \multirow{3}*{GPT-4}  &Vis. Present           &0.2320         &0.1325             &6.9e-322           &True               &0.1280             &0.1370\\
                                &No Vis. Present        &-0.5732        &-0.5047            &3.326E-165         &False              &-0.7088            &-0.3439\\
                                &Both                   &-0.3412        &-0.3722            &3.326E-165         &False              &-0.4906            &-0.2820\\
        \hline
        \multirow{3}*{Gemini}   &Vis. Present           &-0.7057        &-0.6376            &0                  &True               &-0.6434            &-0.6318\\
                                &No Vis. Present        &0.1872         &0.1192             &1.3017E-236        &True               &0.1139             &0.1245\\
                                &Both                   &-0.5185        &-0.5184            &3.326E-165         &False              &-0.6312            &-0.4273\\
        \hline
    \end{tabu}
    \label{tab:genCoefResults}
\end{table*}

\begin{table*}[h]
    \centering
    \caption{Mean coefficients for visualizations (no interactions with tasks).}
    \begin{tabu}{l|S[table-format=1.5]|S[table-format=1.5]|S[table-format=1.5]|S[table-format=1.5]|S[table-format=1.5]|S[table-format=1.7]|S[table-format=1.5]|S[table-format=1.5]|S[table-format=1.4]}
        \hline
        \multirow{2}{*}{\textbf{Visualization}} &\multicolumn{3}{c|}{\textbf{With Visualization}}           &\multicolumn{3}{c|}{\textbf{Without Visualization}}        &\multicolumn{3}{c}{\textbf{Vis. and No Vis.}}\\
        \cline{2-10}
                                                &\textbf{GPT-4}   &\textbf{Gemini}    &\textbf{Both}      &\textbf{GPT-4}   &\textbf{Gemini}    &\textbf{Both}      &\textbf{GPT-4}   &\textbf{Gemini}    &\textbf{Both}\\
        \hline
        Line Chart                              &0.5720             &-0.2104            &0.3615             &0.1672             &-0.5021            &-0.3350            &0.06982            &-0.04325           &0.0266\\
        Bar Chart                               &-0.05956           &-0.1831            &-0.2426            &-0.2468            &0.3080             &0.06120            &-0.3063            &0.1249             &-0.1814\\
        Stacked Bar Chart                       &0.4747             &-0.04813           &0.4266             &0.7244             &-0.6675            &0.05686            &1.199              &-0.7156            &0.4834\\
        100\% Stacked Bar Chart                 &-1.292             &1.342              &0.05057            &-1.131             &0.2363             &-0.8945            &-2.423             &1.579              &-0.8440\\
        Pie Chart                               &-3.104             &1.842              &-1.262             &0.4845             &-0.4848            &-0.0003242         &-2.620             &1.357              &-1.262\\
        Histogram                               &1.751              &-1.725             &0.02577            &-0.6511            &0.5172             &-0.1339            &1.100              &-1.208             &-0.1082\\
        Scatterplot                             &0.4497             &0.4232             &0.8729             &1.478              &-1.200             &0.2775             &1.927              &-0.7771            &1.150\\
        Area Chart                              &0.6932             &0.6223             &1.316              &1.296              &-0.9910            &0.3051             &1.989              &-0.3686            &1.621\\
        Stacked Area Chart                      &-1.429             &0.9821             &-0.4467            &-0.2481            &0.01151            &-0.2366            &-1.677             &0.9936             &-0.6834\\
        Bubble Chart                            &-1.121             &1.183              &0.06183            &0.5422             &-0.2653            &0.2769             &-0.5791            &0.9178             &0.3387\\
        Choropleth Map                          &1.414              &-2.483             &-1.069             &-0.7474            &1.205              &0.4580             &0.6667             &-1.277             &-0.6106\\
        Treemap                                 &1.784              &-2.383             &-0.5997            &-1.503             &1.282              &-0.2208            &0.2803             &-1.101             &-0.8205\\
        \hline
    \end{tabu}
    \label{tab:vizCoefResults}
\end{table*}

\begin{table*}[!h]
    \centering
    \caption{Mean coefficients for tasks (no interactions with visualizations).}
    \begin{tabu}{l|S[table-format=1.4]|S[table-format=1.4]|S[table-format=1.5]|S[table-format=1.5]|S[table-format=1.4]|S[table-format=1.4]|S[table-format=1.4]|S[table-format=1.4]|S[table-format=1.5]}
        \hline
        \multirow{2}{*}{\textbf{Task}}      &\multicolumn{3}{c|}{\textbf{With Visualization}}           &\multicolumn{3}{c|}{\textbf{Without Visualization}}        &\multicolumn{3}{c}{\textbf{Vis. and No Vis.}}\\
        \cline{2-10}
                                            &\textbf{GPT-4}   &\textbf{Gemini}    &\textbf{Both}      &\textbf{GPT-4}   &\textbf{Gemini}    &\textbf{Both}      &\textbf{GPT-4}   &\textbf{Gemini}    &\textbf{Both}\\
        \hline
        Retrieve Value                      &-0.6479            &0.3991             &-0.2488            &-0.8310            &0.4572             &-0.3738            &-1.479             &0.8563             &-0.6225\\
        Find Extremum                       &-1.973             &1.264              &-0.7095            &0.2259             &-0.3824            &-0.1565            &-1.747             &0.8813             &-0.8660\\
        Determine Range                     &-0.1636            &0.1498             &-0.01380           &-0.2322            &0.2604             &0.02816            &-0.3958            &0.4102             &0.01436\\
        Find Correlation/Trend              &1.006              &0.4849             &1.491              &-0.05497           &-0.3682            &-0.4231            &0.9510             &0.1168             &1.068\\
        Make Comparisons                    &-0.5182            &0.8910             &0.3728             &0.8628             &-0.5369            &0.3259             &0.3446             &0.3541             &0.6987\\
        Find Anomalies                      &-0.4557            &-1.240             &-1.696             &0.1304             &-0.3181            &-0.1877            &-0.3253            &-1.558             &-1.884\\
        Find Clusters                       &-0.5395            &0.3645             &-0.1751            &0.2774             &-0.1889            &0.08852            &-0.2621            &0.1756             &-0.08657\\
        Identify the Hierarchical Structure &3.425              &-2.950             &0.4744             &-0.8831            &1.196              &0.3129             &2.542              &-1.754             &0.7874\\
        \hline
    \end{tabu}
    \label{tab:taskCoefResults}
\end{table*}

\begin{table*}[!h]
    \centering
    \caption{Mean coefficients for visualization and task interactions.}
    \begin{tabu}{l|l|S[table-format=1.4]|S[table-format=1.5]|S[table-format=1.5]|S[table-format=1.5]|S[table-format=1.5]|S[table-format=1.7]|S[table-format=1.5]|S[table-format=1.5]|S[table-format=1.5]}
        \hline
        \multirow{2}{*}{\textbf{Visualization}} &\multirow{2}{*}{\textbf{Task}} &\multicolumn{3}{c|}{\textbf{With Visualization}}           &\multicolumn{3}{c|}{\textbf{Without Visualization}}        &\multicolumn{3}{c}{\textbf{Vis. and No Vis.}}\\
        \cline{3-11}
                                                &                               &\textbf{GPT-4}   &\textbf{Gemini}    &\textbf{Both}      &\textbf{GPT-4}   &\textbf{Gemini}    &\textbf{Both}      &\textbf{GPT-4}   &\textbf{Gemini}    &\textbf{Both}\\
        \hline
        \multirow{5}{*}{Line Chart}             &Retrieve Value                 &0.2513             &0.07837            &0.3297             &1.750              &-0.8244            &0.9253             &2.001              &-0.7460            &1.255\\
                                                &Find Extremum                  &-0.2726            &0.1355             &-0.1372            &0.5291             &0.008700           &0.5378             &0.2565             &0.1442             &0.4007\\
                                                &Determine Range                &-0.9321            &0.6569             &-0.2752            &1.640              &-0.9043            &0.7352             &0.7074             &-0.2474            &0.4600\\
                                                &Find Correlation/Trend         &1.714              &-0.9602            &0.7535             &-4.398             &1.864              &-2.533             &-2.684             &0.9043             &-1.780\\
                                                &Make Comparisons               &-0.1883            &-0.1210            &-0.3093            &-0.02280           &0.02267            &-0.0001279         &-0.2111            &-0.09832           &-0.3094\\
        \hline
        \multirow{4}{*}{Bar Chart}              &Retrieve Value                 &0.6114             &0.4350             &1.046              &1.157              &-0.6107            &0.5463             &1.768              &-0.1756            &1.593\\
                                                &Find Extremum                  &-0.7388            &-1.043             &-1.782             &-2.086             &1.186              &-0.8998            &-2.825             &0.1433             &-2.681\\
                                                &Determine Range                &-0.2416            &0.9048             &0.6631             &0.7888             &-0.3374            &0.4514             &0.5472             &0.5673             &1.115\\
                                                &Make Comparisons               &0.3095             &-0.4801            &-0.1707            &-0.1068            &0.07004            &-0.03676           &0.2027             &-0.4101            &-0.2075\\
        \hline
        \multirow{3}{*}{\makecell[l]{Stacked \\ Bar Chart}}
                                                &Retrieve Value                 &-0.4422            &0.2898             &-0.1524            &0.7087             &-0.05651           &0.6522             &0.2666             &0.2333             &0.4998\\
                                                &Find Extremum                  &0.7675             &-0.4049            &0.3626             &0.6926             &-0.9172            &-0.2245            &1.460              &-1.322             &0.1381\\
                                                &Make Comparisons               &0.1494             &0.06699            &0.2164             &-0.6770            &0.3062             &-0.3709            &-0.5276            &0.3731             &-0.1545\\
        \hline
        \multirow{3}{*}{\makecell[l]{100\% \\ Stacked \\ Bar Chart}} 
                                                &Retrieve Value                 &-1.460             &1.273              &-0.1866            &-0.6047            &-0.08910           &-0.6938            &-2.065             &1.184              &-0.8805\\
                                                &Find Extremum                  &-0.5219            &0.1155             &-0.4064            &-0.7320            &-0.08972           &-0.8217            &-1.254             &0.02582            &-1.228\\
                                                &Make Comparisons               &0.6902             &-0.04662           &0.6436             &0.2059             &0.4151             &0.6210             &0.8961             &0.3685             &1.265\\
        \hline
        \multirow{3}{*}{Pie Chart}              &Retrieve Value                 &-0.5742            &0.7811             &0.2069             &-1.328             &0.1364             &-1.192             &-1.902             &0.9175             &-0.9848\\
                                                &Find Extremum                  &-1.532             &0.1133             &-1.419             &3.874              &-0.4027            &3.472              &2.342              &-0.2894            &2.053\\
                                                &Make Comparisons               &-0.9981            &0.9479             &-0.05018           &-2.062             &-0.2186            &-2.280             &-3.060             &0.7293             &-2.330\\
        \hline
        \multirow{3}{*}{Histogram}              &Retrieve Value                 &-1.043             &1.535              &0.4922             &1.778              &-1.079             &0.6996             &0.7351             &0.4568             &1.192\\
                                                &Find Extremum                  &0.7897             &0.3638             &1.153              &-1.732             &0.8630             &-0.8693            &-0.9427            &1.227              &0.2841\\
                                                &Make Comparisons               &2.004              &-3.624             &-1.620             &-0.6971            &0.7328             &0.03577            &1.307              &-2.892             &-1.584\\
        \hline
        \multirow{7}{*}{Scatterplot}            &Retrieve Value                 &-0.4533            &0.1311             &-0.3222            &-0.02240           &0.4353             &0.4129             &-0.4757            &0.5664             &0.09074\\
                                                &Find Extremum                  &-1.351             &4.239              &2.888              &1.322              &-1.416             &-0.09457           &-0.02980           &2.823              &2.793\\
                                                &Determine Range                &-0.2279            &0.3867             &0.1588             &-0.2784            &0.3295             &0.05114            &-0.5063            &0.7162             &0.2100\\
                                                &Find Anomalies                 &-2.769             &-0.7331            &-3.502             &1.015              &-1.246             &-0.2312            &-1.753             &-1.980             &-3.733\\
                                                &Find Clusters                  &-1.649             &1.012              &-0.6365            &0.03590            &-0.09044           &-0.05455           &-1.613             &0.9219             &-0.6910\\
                                                &Find Correlation/Trend         &5.659              &-7.353             &-1.694             &-2.320             &2.729              &0.4091             &3.339              &-4.624             &-1.285\\
                                                &Make Comparisons               &1.240              &2.740              &3.981              &1.726              &-1.941             &-0.2153            &2.966              &0.7990             &3.765\\
        \hline
        \multirow{4}{*}{Area Chart}             &Retrieve Value                 &-0.1458            &-0.7516            &-0.8974            &-0.04543           &0.2074             &0.1620             &-0.1912            &-0.5442            &-0.7354\\
                                                &Find Extremum                  &0.9543             &-0.7363            &0.2181             &0.7891             &-0.7755            &0.01367            &0.1789             &0.05287            &0.2317\\
                                                &Determine Range                &-0.7082            &-0.5284            &-1.237             &-0.6071            &0.4795             &-0.1276            &-1.315             &-0.04894           &-1.364\\
                                                &Find Correlation/Trend         &0.5929             &2.639              &3.232              &2.724              &-2.467             &0.2570             &3.317              &0.1717             &3.489\\
        \hline
        \multirow{4}{*}{\makecell[l]{Stacked \\ Area Chart}}
                                                &Retrieve Value                 &1.847              &-0.7969            &1.050              &1.334              &-0.6257            &0.7088             &3.181              &-1.423             &1.759\\
                                                &Find Extremum                  &-0.4148            &-1.832             &-2.247             &-1.575             &1.426              &-0.1489            &-1.990             &-0.4064            &-2.396\\
                                                &Find Correlation/Trend         &-2.671             &1.237              &-1.434             &-1.901             &0.8875             &-1.014             &-4.572             &2.124              &-2.448\\
                                                &Make Comparisons               &-0.1896            &2.375              &2.185              &1.893              &-1.676             &0.2172             &1.704              &0.6983             &2.402\\
        \hline                            
        \multirow{7}{*}{\makecell[l]{Bubble \\ Chart}}
                                                &Retrieve Value                 &2.308              &-1.383             &0.9247             &-3.707             &1.273              &-2.434             &-1.399             &-0.1108            &-1.509\\
                                                &Find Extremum                  &-1.897             &0.3792             &-1.518             &1.220              &-0.8080            &0.4123             &-0.6766            &-0.4288            &-1.105\\
                                                &Determine Range                &1.946              &-1.270             &0.6761             &-1.775             &0.6931             &-1.082             &0.1712             &-0.5771            &-0.4059\\
                                                &Find Anomalies                 &2.313              &-0.5071            &1.806              &-0.8849            &0.9284             &0.04351            &1.428              &0.4213             &1.849\\
                                                &Find Clusters                  &1.109              &-0.6479            &0.4614             &0.2415             &-0.09844           &0.1431             &1.351              &-0.7463            &0.6045\\
                                                &Find Correlation/Trend         &-4.289             &4.923              &0.6339             &5.840              &-3.382             &2.458              &1.551              &1.540              &3.092\\
                                                &Make Comparisons               &-2.612             &-0.3102            &-2.923             &-0.3929            &1.129              &0.7365             &-3.005             &0.8191             &-2.186\\
        \hline
        \multirow{3}{*}{\makecell[l]{Choropleth \\ Map}}
                                                &Retrieve Value                 &-1.547             &-1.193             &-2.740             &-1.852             &1.690              &-0.1613            &-3.399             &0.4972             &-2.901\\
                                                &Find Extremum                  &0.8702             &1.963              &2.834              &1.635              &-1.435             &0.2001             &2.505              &0.5286             &3.034\\
                                                &Make Comparison                &2.091              &-3.253             &-1.162             &-0.5306            &0.9498             &0.4193             &1.560              &-2.303             &-0.7429\\
        \hline
        \multirow{3}{*}{Treemap}                &Find Extremum                  &1.373              &-2.030             &-0.6562            &-2.146             &0.4131             &-1.733             &-0.7729            &-1.616             &-2.389\\
                                                &Make Comparison                &-3.014             &2.597              &-0.4178            &1.526              &-0.3267            &1.199              &-1.488             &2.270              &0.7816\\
                                                &Identify the Hierarchical Structure
                                                                                &3.425              &-2.950             &0.4744             &-0.8831            &1.196              &0.3129             &2.542              &-1.754             &0.7874\\
        \hline
    \end{tabu}
    \label{tab:vizTaskCoefResults}
\end{table*}

\begin{table*}[!h]
    \centering
    \caption{Two-sided statistically significant results for proportion differences between GPT-4 and Gemini when the visualization is present.  \textbf{Beta Diff?} refers to whether two data sets were beta distributed or not.  If true, the beta-difference distribution\cite{pham1993bayesian} is used, while if it is false, the Wilcoxon signed-rank test is used.  Note: when the Wilcoxon signed-rank test is used, the bounds are estimated from the ECDF.}
    \begin{tabu}{l|l|l|S[table-format=1.7e3]|S[table-format=1.5e1]|S[table-format=1.5e1]}
        \hline
        \multirow{2}*{\textbf{Visualization}}   &\multirow{2}*{\textbf{Task}}   &\multirow{2}*{\textbf{Beta Diff?}} &\multicolumn{3}{c}{\textbf{Two-sided Tests}}\\
        \cline{4-6}
                                                &                               &                                   &\textbf{P-Value}   &\textbf{95\% LB}  &\textbf{95\% UB}\\
        \hline
        Line Chart                              &Retrieve Value                 &True                               &5.239E-07          &0.1915            &0.4214\\
        Line Chart                              &Find Extremum                  &False                              &3.326E-165         &-0.2774           &-0.1253\\
        Line Chart                              &Find Correlation/Trend         &True                               &9.185E-14          &0.3407            &0.5568\\
        Bar Chart                               &Retrieve Value                 &True                               &0.01163            &-0.2821           &-0.03601\\
        Bar Chart                               &Find Extremum                  &False                              &3.326E-165         &-0.02832          &-1.425E-07\\
        Bar Chart                               &Determine Range                &True                               &1.077E-09          &-0.5069           &-0.2720\\
        Stacked Bar Chart                       &Retrieve Value                 &True                               &0.002493           &-0.1758           &-0.03787\\
        Stacked Bar Chart                       &Find Extremum                  &False                              &3.366E-165         &0.03663           &0.2010\\
        Stacked Bar Chart                       &Make Comparisons               &True                               &1.944E-07          &0.1383            &0.3004\\
        100\% Stacked Bar Chart                 &Retrieve Value                 &False                              &3.326E-165         &-0.8624           &-0.7170\\
        100\% Stacked Bar Chart                 &Find Extremum                  &False                              &3.326E-165         &-0.2789           &-0.1392\\
        100\% Stacked Bar Chart                 &Make Comparisons               &False                              &3.326E-165         &-0.8895           &-0.7498\\
        Pie Chart                               &Retrieve Value                 &False                              &3.326E-165         &-0.4454           &-0.2743\\
        Pie Chart                               &Find Extremum                  &False                              &3.326E-165         &-0.3908           &-0.2277\\
        Pie Chart                               &Make Comparisons               &False                              &3.326E-165         &-0.5271           &-0.3485\\
        Histogram                               &Retrieve Value                 &True                               &0.0006654          &0.08430           &0.3088\\
        Histogram                               &Make Comparisons               &False                              &3.326E-165         &0.8306            &0.9464\\
        Scatterplot                             &Retrieve Value                 &True                               &1.528E-06          &-0.4022           &-0.1751\\
        Scatterplot                             &Find Extremum                  &False                              &3.326E-165         &-0.2174           &-0.08750\\
        Scatterplot                             &Determine Range                &True                               &0.04638            &0.001642          &0.2033\\
        Scatterplot                             &Find Anomalies                 &False                              &3.326E-165         &9.758E-08         &1.0912E-05\\
        Scatterplot                             &Find Clusters                  &True                               &0                  &-0.5423           &-0.3415\\
        Scatterplot                             &Find Correlation/Trend         &False                              &3.326E-165         &0.9999            &1.000\\
        Scatterplot                             &Make Comparisons               &False                              &3.326E-165         &4.140E-07         &1.577E-05\\
        Bubble Chart                            &Retrieve Value                 &False                              &3.326E-165         &-0.2047           &-0.07043\\
        Bubble Chart                            &Find Extremum                  &False                              &3.326E-165         &-0.1667           &-0.05833\\
        Bubble Chart                            &Find Anomalies                 &False                              &3.326E-165         &0.3630            &0.5635\\
        Bubble Chart                            &Find Correlation/Trend         &False                              &3.326E-165         &-0.1237           &-0.03287\\
        Bubble Chart                            &Make Comparisons               &False                              &3.326E-165         &-0.1485           &-0.04386\\
        Area Chart                              &Retrieve Value                 &True                               &0.006320           &0.04034           &0.2424\\
        Area Chart                              &Find Extremum                  &True                               &0.01039            &-0.2759           &-0.03732\\
        Area Chart                              &Determine Range                &True                               &0.02488            &0.01399           &0.2043\\
        Area Chart                              &Find Correlation/Trend         &False                              &3.326E-165         &1.643E-07         &9.945E-06\\
        Stacked Area Chart                      &Find Extremum                  &False                              &3.326E-165         &-0.02648          &-9.496E-08\\
        Stacked Area Chart                      &Find Correlation/Trend         &False                              &3.326E-165         &-0.7985           &-0.6293\\
        Stacked Area Chart                      &Make Comparisons               &False                              &3.326E-165         &-0.2991           &-0.1866\\
        Choropleth Map                          &Retrieve Value                 &False                              &2.642E-162         &-2.721E-06        &-1.049E-08\\
        Choropleth Map                          &Find Extremum                  &False                              &3.326E-165         &0.1198            &0.3121\\
        Choropleth Map                          &Make Comparisons               &False                              &3.326E-165         &0.7231            &0.8678\\
        Treemap                                 &Find Extremum                  &False                              &2.150E-84          &-6.782E-06        &0.02747\\
        Treemap                                 &Make Comparisons               &False                              &3.326E-165         &-0.5927           &-0.4028\\
        Treemap                                 &Identify Hierarchical Structure&False                              &3.326E-165         &1.000             &1.000\\
        \hline
    \end{tabu}
    \label{tab:llmDiffResults}
\end{table*}

\begin{table*}[!h]
    \centering
    \caption{Statistically significant results for proportion differences between visualization and no visualization present.  \textbf{Beta Diff?} refers to whether two data sets were beta distributed or not.  If true, the beta-difference distribution\cite{pham1993bayesian} is used, while if it is false, the Wilcoxon signed-rank test is used.  Note that when the Wilcoxon signed-rank test is used, the bounds are estimated from the ECDF.  Thus, lower bounds calculated from the ECDF less than or equal to zero indicate poor estimations.}
    \begin{tabu}{l|l|l|l|S[table-format=1.6e3]|S[table-format=1.5e1]}
        \hline
        \textbf{LLM}    &\textbf{Visualization} &\textbf{Task}                      &\textbf{Beta Diff?} &\textbf{P-Value}  &\textbf{95\% LB}\\
        \hline
        GPT-4         &Line Chart             &   Retrieve Value                  &True                &0.02390           &0.0209\\
        GPT-4         &Line Chart             &   Find Correlation/Trend          &False               &1.663E-165        &0.8161\\
        GPT-4         &Bar Chart              &   Retrieve Value                  &True                &0.009040          &0.0439\\
        GPT-4         &Stacked Bar Chart      &   Make Comparisons                &True                &0.0006223         &0.06318\\
        GPT-4         &100\% Stacked Bar Chart&   Retrieve Value                  &False               &5.883E-161        &6.330E-08\\
        GPT-4         &Histogram              &   Find Extremum                   &False               &4.067E-165        &0.01524\\
        GPT-4         &Histogram              &   Make Comparisons                &True                &2.782E-19         &0.4590\\
        GPT-4         &Scatterplot            &   Find Correlation/Trend          &False               &2.490E-164        &0.007176\\
        GPT-4         &Scatterplot            &   Make Comparisons                &False               &1.698E-165        &3.450E-07\\
        GPT-4         &Bubble Chart           &   Retrieve Value                  &False               &2.130E-155        &-3.445E-06\\
        GPT-4         &Bubble Chart           &   Determine Range                 &False               &1.663E-165        &0.1679\\
        GPT-4         &Bubble Chart           &   Find Anomalies                  &True                &1.013E-07         &0.2260\\
        GPT-4         &Area Chart             &   Find Correlation/Trend          &False               &1.673E-165        &2.066E-08\\
        GPT-4         &Stacked Area Chart     &   Find Correlation/Trend          &False               &1.415E-92         &-2.793E-06\\
        GPT-4         &Choropleth Map         &   Find Extremum                   &False               &3.172E-46         &-0.03726\\
        GPT-4         &Choropleth Map         &   Make Comparisons                &True                &0.002537          &0.06992\\
        GPT-4         &Treemap                &   Find Extremum                   &False               &6.947E-92         &-3.216E-06\\
        GPT-4         &Treemap                &   Identify Hierarchical Structure &False               &1.663E-165        &0.1870\\
        Gemini          &Line Chart             &   Find Correlation/Trend          &False               &1.663E-165        &0.2841\\
        Gemini          &Stacked Bar Chart      &   Find Extremum                   &False               &1.996E-149        &-0.006828\\
        Gemini          &Stacked Bar Chart      &   Make Comparisons                &True                &3.491E-22         &0.3535\\
        Gemini          &100\% Stacked Bar Chart&   Retrieve Value                  &True                &2.598E-25         &0.5573\\
        Gemini          &100\% Stacked Bar Chart&   Find Extremum                   &False               &1.663E-165        &0.1212\\
        Gemini          &100\% Stacked Bar Chart&   Make Comparisons                &False               &1.663E-165        &0.05043\\
        Gemini          &Pie Chart              &   Retrieve Value                  &False               &1.663E-165        &0.2214\\
        Gemini          &Pie Chart              &   Make Comparisons                &False               &1.663E-165        &0.3445\\
        Gemini          &Scatterplot            &   Find Extremum                   &False               &1.663E-165        &0.1472\\
        Gemini          &Scatterplot            &   Determine Range                 &True                &8.863E-07         &0.2045\\
        Gemini          &Scatterplot            &   Find Clusters                   &True                &1.787E-11         &0.3016\\
        Gemini          &Scatterplot            &   Make Comparisons                &False               &1.663E-165        &0.2000\\
        Gemini          &Bubble Chart           &   Retrieve Value                  &False               &1.997E-165        &0.02609\\
        Gemini          &Bubble Chart           &   Find Extremum                   &False               &2.127E-150        &-0.007374\\
        Gemini          &Bubble Chart           &   Find Correlation/Trend          &False               &2.279E-165        &0.008458\\
        Gemini          &Area Chart             &   Find Extremum                   &True                &1.208E-08         &0.2551\\
        Gemini          &Area Chart             &   Find Correlation/Trend          &False               &1.663E-165        &0.2523\\
        Gemini          &Stacked Area Chart     &   Find Correlation/Trend          &True                &1.075E-18         &0.4594\\
        Gemini          &Stacked Area Chart     &   Make Comparisons                &False               &1.663E-165        &0.2373\\
        Gemini          &Choropleth Map         &   Find Extremum                   &True                &9.811E-05         &0.1360\\
        \hline
    \end{tabu}
    \label{tab:vizDiffResults}
\end{table*}

\end{document}